\documentclass[pra,twocolumn,amsmath,amssymb,superscriptaddress]{revtex4-1}
\usepackage{epsfig,amsmath}
\usepackage{subfigure}
\usepackage{graphicx}
\usepackage{dcolumn}
\usepackage{stmaryrd}
\usepackage{mathrsfs}
\usepackage{pifont}
\usepackage{amsthm}
\usepackage{amssymb}
\usepackage{bm}
\usepackage{latexsym}
\usepackage[colorlinks=true,linkcolor=blue,citecolor=blue]{hyperref}
\usepackage{color}
\usepackage{epstopdf}

\usepackage{booktabs}
\usepackage{threeparttable}
\usepackage{multirow}
\usepackage{array}
\usepackage{mathdots}

\newcommand{\red}[1]{\textcolor{red}{#1}}

\begin{document}

\title{From Liouville equation to universal quantum control: A study of generating ultra highly squeezed states}

\author{Zhu-yao Jin}
\affiliation{School of Physics, Zhejiang University, Hangzhou 310027, Zhejiang, China}

\author{J. Q. You}
\affiliation{School of Physics, Zhejiang University, Hangzhou 310027, Zhejiang, China}

\author{Jun Jing}
\email{Contact author: jingjun@zju.edu.cn}
\affiliation{School of Physics, Zhejiang University, Hangzhou 310027, Zhejiang, China}

\date{\today}

\begin{abstract}
We find that the seemingly disparate control approaches for classical and quantum continuous-variable systems can be unified via differential manifolds of the ancillary representations. For classical systems, the ancillary representation is defined by the time-dependent ancillary canonical variables resulting from symplectic transformation over the original canonical variables. Under the Hamilton-Jacobi theory, the ancillary canonical variables act as dynamical invariants to guide the system nonadiabatically through the entire phase space. The second quantization of the Liouville equation for the dynamical invariants leads to the Heisenberg equation for the relevant ancillary operators, which is found to be a sufficient condition to activate nonadiabatic passages towards arbitrary target states in both Hermitian and non-Hermitian systems and yield constrained exact solutions of the time-dependent Schr\"odinger equation. Using the non-Hermitian Hamiltonian rigorously derived from the Lindblad master equation, our theory is exemplified by the generation of single-mode squeezed states with a squeezing level about $29.3$ dB and double-mode squeezed states with $20.6$ dB, respectively.
\end{abstract}

\maketitle

\section{Introduction}

In classical mechanics, the system dynamics can be described by the dynamical invariant extracted from the Hamilton-Jacobi theory~\cite{Arnol2013mathematical,Landau2013course}. In quantum mechanics, the exact solution of time-dependent Schr\"odinger equation~\cite{Schrodinger1926Undulatory} requires mutual commutation of the Hamiltonian at distinct moments. Otherwise, the dynamics in general admits only formal solution, e.g., the Dyson series~\cite{Dyson1949TheRadiation} and the Magnus expansion~\cite{Blanes2009Magnus}, or resorts to the effective Hamiltonian~\cite{Nakajima1955Perturbation,Schrieffer1966Relation,Morris1983Reduction,James2007Effective}. Open-quantum-system dynamics appears to further widen the discrepancy between classical and quantum mechanics~\cite{Carmichael1999statistical}.

Kinematics draw parallels between classical and quantum mechanics in both closed and open scenario~\cite{Ashida2020NonHermitian}. For instance, in classical shortcut-to-adiabaticity (STA) protocols~\cite{Jarzynski2013Generating,Deffner2014Classical,Jarzynski2017Fast}, an ancillary Hamiltonian drives a harmonic oscillator to rapidly go through the entire phase space by following the adiabatic invariant of the system Hamiltonian. For closed quantum systems, the nonadiabatic dynamics of a harmonic oscillator~\cite{Lewis1969Exact} and a two-level system~\cite{Chen2011Lewis} can be formulated by the Lewis-Riesenfeld invariant. This invariant can be applied to a few open systems described by the non-Hermitian Hamiltonian~\cite{Gao1992Invariant,Ibanez2011Shortcuts,Luo2015Dynamical}. By virtue of the parity-time symmetry~\cite{Bender2007Faster} that provides a broader criteria for ensuring a real eigenspectrum, the dynamics of an open two-mode system~\cite{Jan2019Nonclassical} can be manipulated around the exceptional points~\cite{Friederike2025Crossing} of the Hamiltonian. Spectrum-based controls of non-Hermitian systems~\cite{Metelmann2015Nonreciprocal} are limited in scalability~\cite{Liu2023Practical,Zhang2025Exponential}, lose track of the global phase accumulated during the evolution, and enforce artificial state renormalization. Often they demand stringent premises in experiments, such as a squeezing level as high as $20\sim30$ dB for a single-mode system~\cite{Chen2021Shortcuts}.

Among nonclassical states~\cite{Roman2017Squeezed,Gerry2023Introductory}, squeezed states are particularly fruitful and experimentally viable resources for continuous-variable quantum information processing, including quantum teleportation~\cite{Braunstein1998Teleportation}, quantum metrology~\cite{Mamaev2025NonGaussion}, quantum dense coding~\cite{Braunstein2000Dense}, detection of gravitational waves~\cite{Roman2017Squeezed}, fault-tolerant quantum computation (FTQC)~\cite{Menisucci2014Fault}, and the Gottesman-Kitaev-Preskill (GKP) state preparation~\cite{Fluhmann2019Encoding}. The utility of squeezed states depends critically on the squeezing level, with thresholds of $10$ dB for high-fidelity quantum teleportation~\cite{He2024Performance}, and $12.7$ dB and $20.5$ dB for GKP-based~\cite{Larsen2021Fault} and cluster-state-based~\cite{Menisucci2014Fault} FTQC, respectively. However, the achievable squeezing is severely limited by dissipation~\cite{Miifmmode2012Noiseless}. To our best knowledge, the highest squeezing level reaches about $15$ dB for single-mode states both theoretically~\cite{Korobko2023Fundamental} and experimentally~\cite{Vahlbruch2016Detection,Cai2025Quantum}; and for two-mode squeezing, the records are about $15$ dB in theory~\cite{Sutherland2021Universal,Superposition2021Cardoso} and $10$ dB in experiment~\cite{Eberle2013Stable}.

In this work, we develop an operational framework from a geometric perspective to unify the dynamical control over classical and quantum continuous-variable systems under time-dependent Hamiltonian. It establishes, for the first time, a systematic connection between quantum control for Hermitian and non-Hermitian systems and analytical mechanics as a synthesis of canonical transformation, Hamilton-Jacobi theory, Liouville equation, dynamical invariant, and symplectic transformation. Our theory is fundamentally built on the dialogue between the generating function for both original and ancillary canonical variables and the gauge potential for ancillary operators. In particular, the nonadiabatic dynamics of classical systems is formulated in terms of time-dependent canonical variables by a symplectic transformation determined by the generating function. The transformed Hamiltonian satisfies the Hamilton-Jacobi equation~\cite{Arnol2013mathematical,Landau2013course} if and only if the ancillary variables satisfy the Liouville equation with the original Hamiltonian. In this case, the ancillary variables become dynamical invariants governing the quantum trajectory. The Liouville equation is mapped to the Heisenberg equation for time-dependent ancillary bosonic operators up to a gauge field on the standard second quantization~\cite{Dirac1925Fundamental}. The ancillary operators are related to the original bosonic operators through the same symplectic transformation matrix as its classical counterpart, establishing a manifold structure captured by the gauge field. In both Hermitian and non-Hermitian systems, we prove that the Heisenberg equation for the ancillary operators constitutes a sufficient condition for both constrained exact solutions to the time-dependent Schr\"odinger equation and system controllability. As a paradigmatic example of running our framework, we propose to generate both single-mode and two-mode squeezed states of much higher squeezing levels than the existing protocols, in which we employ the non-Hermitian Hamiltonian rigorously derived from the adjoint Lindblad master equation~\cite{Metelmann2015Nonreciprocal,Wang2019Nonreciprocity}.

The rest part of the main text is organized as follows. In Sec.~\ref{generalClas}, a universal classical control framework is introduced to construct the Liouville equation for the dynamical invariants. In Sec.~\ref{generalQuant}, the Liouville equation is mapped to the Heisenberg equation for the ancillary operators under quantization, which applies to both Hermitian and non-Hermitian Hamiltonian. In Sec.~\ref{SingSquez}, our theory is exemplified by generating an ultra-highly squeezed state in a single bosonic-mode system under a time-dependent non-Hermitian Hamiltonian. Its fault-tolerance against the parametric fluctuation is found to be sustained by an error-correction mechanism inherently within our protocol. In Sec.~\ref{DoubleSquez}, we discuss the generation of an ultra-highly squeezed double-mode state. The entire work is summarized in Sec.~\ref{Conclusion}.

Appendix~\ref{UCCone} constructs the dynamical invariants of time-dependent single- and double-mode classical systems by canonical transformation and Hamilton-Jacobi theory. For general $N$-mode systems, Appendix~\ref{DervonNeuClas} details the derivation from the Hamilton-Jacobi equation to the Liouville equation. Appendix~\ref{HJscale} gives an example about using the Hamilton-Jacobi equation in Appendix~\ref{UCCone} to control a classical scale-invariant system. Appendices~\ref{quadratic} and~\ref{nonlinear} prove that the ancillary operators constrained by the Heisenberg equation can activate nonadiabatic passages for the systems governed by quadratic and nonlinear squeezing Hamiltonian, respectively. Appendix~\ref{SufNecDiag} proves that the Heisenberg equation with non-Hermitian Hamiltonian renders the dynamical invariants in either bra or ket space under the biorthogonal assumption. Appendices~\ref{HeffDerivesingle} and \ref{HeffDerivedouble} demonstrate how to exactly attain the non-Hermitian squeezing Hamiltonian from the Lindblad master equation without ignoring quantum jump terms for single- and double-mode systems, respectively. Appendix~\ref{recipe} presents the completed symplectic transformation matrix for the two-mode squeezing system, and provides a brief recipe for the general ancillary operators of $N$-mode bosonic systems with squeezing interactions.

\section{Universal control}

\subsection{Universal framework for classical control}\label{generalClas}

Consider a general system composed of $N$ classical harmonic oscillators or resonators, governed by the time-dependent Hamiltonian $H(\vec{q},\vec{p},t)$, where $\vec{q}=(q_1,q_2,\cdots,q_N)$ and $\vec{p}=(p_1,p_2,\cdots,p_N)$ denote the canonical coordinates and momenta, respectively. This system can be equivalently described by the complex canonical variables $\vec{a}=(a_1,a_2,\cdots,a_N)$ and $\vec{a}^*=(a_1^*,a_2^*,\cdots,a_N^*)$, where~\cite{STROCCHI1966Complex}
\begin{equation}\label{CanoTrans}
a_k\equiv\frac{1}{\sqrt{2}}\left(\sqrt{m_k\omega_k}q_k+\frac{i}{\sqrt{m_k\omega_k}}p_k\right),\quad 1\leq k\leq N,
\end{equation}
with the mass $m_k$ of the $k$th harmonic oscillator and its frequency $\omega_k$. These canonical variables satisfy the Poisson bracket $\{a_j,a_k^*\}_{\vec{q},\vec{p}}=-i\{a_j,a_k^*\}_{\vec{a},\vec{a}^*}=-i\delta_{jk}$, where $\{A,B\}_{\vec{a},\vec{a}^*}\equiv\sum_{k=1}^N(\partial_{a_k}A\partial_{\partial a_k^*}B-\partial_{a_k^*}A\partial_{a_k}B)$ with arbitrary variables $A$ and $B$. The Hamilton's canonical equations of $a_k$ and $a_k^*$ are~\cite{STROCCHI1966Complex}
\begin{equation}\label{HamEq}
i\frac{da_k}{dt}=\frac{\partial H(\vec{a},\vec{a}^*,t)}{\partial a_k^*},\quad i\frac{da_k^*}{dt}=-\frac{\partial H(\vec{a},\vec{a}^*,t)}{\partial a_k}.
\end{equation}
They typically yield a set of coupled differential equations and thus hard to find an analytical solution unless obtaining a time-dependent invariant manifold. Particularly, one can pose a time-dependent canonical transformation
\begin{equation}\label{TransR}
\left[\vec{u}(\vec{a},\vec{a}^*,t), \vec{u}^*(\vec{a},\vec{a}^*,t)\right]^T=\mathcal{M}(t)(\vec{a}, \vec{a}^*)^T,
\end{equation}
where $\mathcal{M}(t)$ is a time-dependent $2N\times2N$ symplectic matrix and the superscript $T$ denotes the transposition that converts a row vector into a column vector. The ancillary canonical variables $\vec{u}(\vec{a},\vec{a}^*,t)$ and $\vec{u}^*(\vec{a},\vec{a}^*,t)$ are explicitly time-dependent and they satisfy the relation $\{u_j,u_k^*\}_{\vec{q},\vec{p}}=-i\{u_j,u_k^*\}_{\vec{u},\vec{u}^*}=-i\delta_{jk}$.

Under the symplectic structure formulated by Eq.~(\ref{TransR}), the transformed Hamiltonian takes the form of~\cite{Arnol2013mathematical,Landau2013course,STROCCHI1966Complex}
\begin{equation}\label{HamKGen}
K(\vec{u},\vec{u}^*,t)=H(\vec{a},\vec{a}^*,t)+\frac{\partial F_3(\vec{a},\vec{u}^*,t)}{\partial t},
\end{equation}
with the type-III generating function $F_3(\vec{a},\vec{u}^*,t)$ constrained by the partial derivatives $u_k=-i\partial_{u_k^*}F_3(\vec{a},\vec{u}^*,t)$ and $a_k^*=-i\partial_{a_k} F_3(\vec{a},\vec{u}^*,t)$. Equation~(\ref{HamKGen}) reduces to the Hamilton-Jacobi equation when $K(\vec{u},\vec{u}^*,t)=0$. In this case, $u_k$ and $u_k^*$ with $1\leq k\leq N$ become dynamical invariants since their Hamilton's canonical equations are
\begin{equation}\label{HamKEq}
\frac{du_k}{dt}=-i\frac{\partial K(\vec{u},\vec{u}^*,t)}{\partial u_k^*}=0,\quad \frac{du_k^*}{dt}=i\frac{\partial K(\vec{u},\vec{u}^*,t)}{\partial u_k}=0.
\end{equation}
In Appendix~\ref{UCCone}, we demonstrate the construction of the dynamical invariants for time-modulating classical single- and double-mode systems by properly setting the parameters in the original Hamiltonian $H(\vec{a},\vec{a}^*,t)$. Consequently, the system trajectories in the complex phase space $(a_k,a_k^*)$ as well as the exact solutions to the Hamilton's canonical equations~(\ref{HamEq}) can be obtained by the inverse transformation of Eq.~(\ref{TransR}).

More importantly, the explicit time dependence of $u_k(\vec{a},\vec{a}^*,t)$ and $u_k^*(\vec{a},\vec{a}^*,t)$ in Eq.~(\ref{HamKEq}) leads to the Liouville equation (see Appendix~\ref{DervonNeuClas} for details) as
\begin{equation}\label{vonNeuClas}
\frac{\partial J_k(\vec{a},\vec{a}^*,t)}{\partial t}=-i\left\{H(\vec{a},\vec{a}^*,t),J_k(\vec{a},\vec{a}^*,t)\right\}_{\vec{a},\vec{a}^*},
\end{equation}
with $J_k=u_k$ and $u_k^*$. In Appendix~\ref{HJscale}, it is interesting to find that our general theory based on Eq.~(\ref{vonNeuClas}) suffices to encompass the classical STA protocols~\cite{Jarzynski2013Generating,Deffner2014Classical} for scale-invariant systems~\cite{Deffner2014Classical}.

\subsection{Universal framework for quantum control}\label{generalQuant}

In quantum mechanics, the Liouville equation~(\ref{vonNeuClas}) translates to the Heisenberg equation, by which one can control both Hermitian and non-Hermitian continuous-variable systems. Our study is conducted on a general system consisting of $N$ bosonic modes, described by the annihilation and creation operators, i.e., $\hat{a}_k$ and $\hat{a}_k^\dagger$ with $k=1,2,\cdots,N$. The system dynamics is given by ($\hbar\equiv1$),
\begin{equation}\label{SchTime}
i\frac{\partial}{\partial t}|\psi(t)\rangle=\hat{H}(t)|\psi(t)\rangle,
\end{equation}
where $|\psi(t)\rangle$ is the pure-state solution and the time-dependent Hamiltonian $\hat{H}(t)$ takes the form of
\begin{equation}\label{Ham}
\hat{H}(t)=\begin{pmatrix}\vec{a}^\dagger&\vec{a}\end{pmatrix}H^a(t)
\begin{pmatrix}\vec{a}^T\\(\vec{a}^\dagger)^T\end{pmatrix},
\end{equation}
with the row operator vectors $\vec{a}=(\hat{a}_1,\hat{a}_2,\cdots,\hat{a}_N)$ and $\vec{a}^\dagger\equiv(\hat{a}_1^\dagger,\hat{a}_2^\dagger,\cdots,\hat{a}_N^\dagger)$. The diagonal and off-diagonal elements of $2N\times2N$ coefficient matrix $H^a(t)$ correspond to the eigenfrequencies of the bosonic mode and the squeezing and (or) exchange interactions between them, respectively. More general nonlinear models, e.g., spontaneous parametric down-conversion~\cite{Maria2011Nonlinear,Borshchevskaya2015Three,Akbari2016Third,Chang2020Observation}, can also be addressed by our framework to work out the dynamical invariants (see Appendix~\ref{nonlinear} for details).

The time-dependent Schr\"odinger equation~(\ref{SchTime}) typically admits only formal solutions~\cite{Dyson1949TheRadiation,Blanes2009Magnus} due to the noncommutativity of $\hat{H}(t)$ at distinct moments. Applying the same symplectic transformation $\mathcal{M}(t)$ in Eq.~(\ref{TransR}) as for the classical resonators to the original bosonic operators $\{\hat{a}_k, \hat{a}^{\dagger}_k\}$, one can construct a set of time-dependent ancillary operators $\{\hat{\mu}_k(t), \hat{\mu}^{\dagger}_k(t)\}$:
\begin{equation}\label{TransM}
[\vec{\mu}, \vec{\mu}^\dagger]^T=\mathcal{M}(t)(\vec{a}, \vec{a}^{\dagger})^T,
\end{equation}
where $\vec{\mu}=\vec{\mu}(t)=[\hat{\mu}_1(t),\hat{\mu}_2(t),\cdots,\hat{\mu}_N(t)]$ and $\vec{\mu}^\dagger=\vec{\mu}^\dagger(t)=[\hat{\mu}_1^\dagger(t),\hat{\mu}_2^\dagger(t),\cdots,\hat{\mu}_N^\dagger(t)]$. The canonical commutation relations are preserved by the symplectic structure, i.e., $[\hat{\mu}_k(t),\hat{\mu}_m^\dagger(t)]=\delta_{km}$. Consequently, we have
\begin{equation}\label{HamMu}
\begin{aligned}
\hat{H}(t)&=\begin{pmatrix}\vec{\mu}^\dagger&\vec{\mu}\end{pmatrix}
\left[\mathcal{M}^\dagger(t)\right]^{-1}H^a(t)\mathcal{M}^{-1}(t)
\begin{pmatrix}\vec{\mu}^T\\(\vec{\mu}^\dagger)^T\end{pmatrix}\\
&=\begin{pmatrix}\vec{\mu}^\dagger&\vec{\mu}\end{pmatrix}H^\mu(t)
\begin{pmatrix}\vec{\mu}^T\\(\vec{\mu}^\dagger)^T\end{pmatrix},
\end{aligned}
\end{equation}
where $H^\mu(t)$ is a $2N\times2N$ rotated coefficient matrix.

To proceed, we transform the system dynamics into the stationary ancillary representation via a unitary transformation $\mathcal{V}(t)$, which satisfies
\begin{equation}\label{Unitrans}
\mathcal{V}^\dagger(t)\hat{\mu}_k(t)\mathcal{V}(t)=\hat{\mu}_k(0), \quad k=1,2,\cdots,N,
\end{equation}
where $\{\hat{\mu}_k(0)\}$ are the stationary ancillary operators. $\mathcal{V}(t)$ can be explicitly determined once $\mathcal{M}(t)$ in Eq.~(\ref{TransR}) or (\ref{TransM}) is specified~\cite{Jin2025Universal,Jin2025Entangling,Jin2025UniNon,Jin2026Unibosonic}, see, e.g., Eq.~(\ref{Vone}). In the rotating frame with respect to $\mathcal{V}(t)$, $\hat{H}(t)$ in Eq.~(\ref{HamMu}) is transformed as
\begin{equation}\label{HamMuSta}
\begin{aligned}
\hat{H}_{\rm rot}(t)&=\mathcal{V}^\dagger(t)\hat{H}(t)\mathcal{V}(t)-i\mathcal{V}^\dagger(t)
\frac{\partial\mathcal{V}(t)}{\partial t}\\
&=\begin{pmatrix}\vec{\mu}_0^\dagger&\vec{\mu}_0\end{pmatrix}\Big[H^\mu(t)-\mathcal{A}(t)\Big]
\begin{pmatrix}\vec{\mu}_0^T\\(\vec{\mu}_0^\dagger)^T\end{pmatrix},\\
&\equiv\begin{pmatrix}\vec{\mu}_0^\dagger&\vec{\mu}_0\end{pmatrix}\mathcal{H}(t)
\begin{pmatrix}\vec{\mu}_0^T\\(\vec{\mu}_0^\dagger)^T\end{pmatrix},
\end{aligned}
\end{equation}
where $\vec{\mu}_0\equiv\vec{\mu}(0)=[\hat{\mu}_1(0),\hat{\mu}_2(0),\cdots,\hat{\mu}_N(0)]$, and the gauge or holonomic potential~\cite{Zhang2023Geometric} is given by $\mathcal{A}(t)=i[\mathcal{M}^\dagger(0)]^{-1}\mathcal{M}^{-1}(t)\dot{\mathcal{M}}(t)\mathcal{M}^{-1}(0)$ with the stationary symplectic matrix $\mathcal{M}(0)$.

\emph{Main results.---} Upon the second quantization, i.e., the translation of the classical Hamiltonian to its quantum counterpart $H(\vec{a},\vec{a}^*,t)\mapsto\hat{H}(t)$, the Poisson bracket to the commutator $\{A,B\}_{\vec{a},\vec{a}^*}\mapsto[A,B]$, and the ancillary canonical variables to the time-dependent ancillary bosonic operators $u_k(\vec{a},\vec{a}^*,t)\mapsto\hat{\mu}_k(t)$ and $u_k^*(\vec{a},\vec{a}^*,t)\mapsto\hat{\mu}_k^\dagger(t)$, $k\in\{1,2,\cdots,N\}$, the Liouville equation~(\ref{vonNeuClas}) can be mapped to the Heisenberg equation:
\begin{equation}\label{vonNeu}
\frac{\partial\mathcal{J}_k(t)}{\partial t}=-i\left[\hat{H}(t), \mathcal{J}_k(t)\right],
\end{equation}
where the dynamical invariants $\mathcal{J}_k(t)=\tilde{\mu}_k(t)$ and $\tilde{\mu}_k^\dagger(t)$, are the ancillary operators dressed with a gauge-invariant global phase:
\begin{equation}\label{Quanti}
\hat{\mu}_k(t)\mapsto\tilde{\mu}_k(t)\equiv e^{if_k(t)}\hat{\mu}_k(t), \quad \hat{\mu}_k^\dagger(t)\mapsto\tilde{\mu}_k^\dagger(t),
\end{equation}
with $f_k(t)\equiv\int_0^t\mathcal{H}_{kk}(s)ds$. In Appendix~\ref{quadratic}, we prove that for a closed $N$-mode bosonic system with squeezing Hamiltonian, Eq.~(\ref{vonNeu}) is a necessary and sufficient condition for the complete or partial diagonalization of the coefficient matrix $\mathcal{H}(t)$ in Eq.~(\ref{HamMuSta}). A partially diagonalized coefficient matrix suffices to yield the dynamics of $\hat{\mu}_k(0)$ and $\hat{\mu}_k^\dagger(0)$ in the original picture:
\begin{equation}\label{HermDynOri}
\hat{\mu}_k(0)\rightarrow e^{-if_k(t)}\hat{\mu}_k(t),\quad \hat{\mu}_k^\dagger(0)\rightarrow e^{if_k(t)}\hat{\mu}_k^\dagger(t).
\end{equation}
Also it serves as the constrained exact solution to the time-dependent Schr\"odinger equation~(\ref{SchTime}) when the system state is initially prepared by $\hat{\mu}_k(0)$ or $\hat{\mu}_k^\dagger(0)$.

\emph{Non-Hermitian case.---} In practical scenarios, quantum systems are inevitably under the influence of the external noises. In comparison to the Lindblad master equation~\cite{Carmichael1999statistical}, the Schr\"odinger equation with non-Hermitian Hamiltonian is scalable in numerical simulation~\cite{Metelmann2015Nonreciprocal,Wang2019Nonreciprocity}. Moreover, our dynamical-invariant-based control and the associated Heisenberg equation~(\ref{vonNeu}) also apply to non-Hermitian Hamiltonian, i.e., $\hat{H}(t)\neq \hat{H}^\dagger(t)$. Under the biorthogonal assumption~\cite{Brody2013Biorhogonal}, the system dynamics can be described by two sets of time-dependent Schr\"odinger equations as
\begin{equation}\label{Sch}
i\frac{\partial}{\partial t}|\psi(t)\rangle=\hat{H}(t)|\psi(t)\rangle,\quad i\frac{\partial}{\partial t}|\phi(t)\rangle=\hat{H}^\dagger(t)|\phi(t)\rangle
\end{equation}
with $|\psi(t)\rangle$ and $\langle\phi(t)|$ describing the pure-state solutions in the ket and bra spaces, respectively. In Appendix~\ref{SufNecDiag}, we prove that the nonunitary dynamics of the system can be described by
\begin{equation}\label{NonHermDynOri}
\hat{\mu}_k^\dagger(0)\rightarrow e^{if_k(t)}\hat{\mu}_k^\dagger(t)
\end{equation}
when focusing on Eq.~(\ref{vonNeu}) in the ket space with $\mathcal{J}_k(t)=\tilde{\mu}_k^\dagger(t)$. In this case, the rotated Hamiltonian $\hat{H}_{\rm rot}(t)$ takes the same form as Eq.~(\ref{HamMuSta}) yet with a non-Hermitian coefficient matrix $\mathcal{H}(t)$, where all the elements in the $(N+k)$th row and $k$th column are zero. In contrast to the Hermitian case, here the global phase $f_k(t)$ is generally a complex number, the imaginary part of which indicates the probability nonconservation in open quantum systems. We can, however, always have a real $f_k(\tau)$ at the end of control, as demonstrated in the following examples.

\section{Generation of single-mode squeezed state}\label{SingSquez}

In this section, we concentrate on the non-Hermitian passages of a SU(1,1) generator by applying the general theory in Sec.~\ref{generalQuant} to the creation of a highly squeezed state for a single-mode system, given that we have discussed various Hermitian passages in Refs.~\cite{Jin2025Universal,Jin2025Entangling,Jin2026Unibosonic} and the non-Hermitian passages of a SU(2) generator in Ref.~\cite{Jin2025UniNon}. The single-mode squeezed vacuum state is defined as~\cite{Yuen1976Twophoton,Fabre2020Modes,Gerry2023Introductory,Eriksson2024Universal,Cai2025Quantum}
\begin{equation}\label{SquezSing}
\begin{aligned}
|\xi\rangle&=S_1(\xi)|0\rangle=e^{\frac{1}{2}\left[\xi^*\hat{a}_1^2-\xi(\hat{a}_1^\dagger)^2\right]}|0\rangle\\
&=\frac{1}{\sqrt{\cosh r}}\sum_{n=0}^\infty\frac{\sqrt{(2n)!}}{2^nn!}(-e^{-i\phi}\tanh r)^n|2n\rangle,
\end{aligned}
\end{equation}
where the squeezing operator $S_1(\xi)$ is characterized by $\xi=r\exp(-i\phi)$ with strength $r$ and phase $\phi$. The dynamics of a dissipative single-mode system driven by a two-photon driving field can be described by the Lindblad master equation~\cite{Carmichael1999statistical}. In the Heisenberg picture, such open-quantum-system dynamics can be rigorously mapped to the time-dependent Schr\"odinger equation with a non-Hermitian squeezing Hamiltonian (see Appendix~\ref{HeffDeriveSing} for details):
\begin{equation}\label{HamNHsing}
\hat{H}(t)=\left[\omega(t)-i\frac{\gamma}{2}\right]\hat{a}_1^\dagger\hat{a}_1
+\left[\Omega(t)e^{i\varphi}\left(\hat{a}_1^\dagger\right)^2+{\rm H.c.}\right],
\end{equation}
where $\omega(t)$, $\Omega(t)$, $\varphi$, and $\gamma$ are the mode eigenfrequency, the two-photon driving intensity, the driving phase, and the decay rate, respectively. In circuit-QED systems, the cavity loss effect $\gamma>0$ can be effectively converted to the gain effect $\gamma<0$~\cite{Fernando2018PT,Purkayastha2020Emergent} by coupling to a gain medium~\cite{Liu2014Photon,Liu2015Semiconductor,Stehlik2016Double}.

For a single-mode system, Eq.~(\ref{TransM}) can be written as $[\hat{\mu}(t), \hat{\mu}^\dagger(t)]^T=\mathcal{M}(t)(\hat{a}_1, \hat{a}_1^\dagger)^T$, where the symplectic matrix is explicitly chosen as
\begin{equation}\label{Mone}
\mathcal{M}(t)=\begin{pmatrix}\cosh\theta(t)&-\sinh\theta(t)e^{-i\alpha(t)}
\\-\sinh\theta(t)e^{i\alpha(t)}&\cosh\theta(t)\end{pmatrix}.
\end{equation}
Here the time-dependent parameters $\theta(t)$ and $\alpha(t)$ are associated with the squeezing strength and phase, respectively. Consequently, the unitary transformation $\mathcal{V}(t)$ in Eq.~(\ref{Unitrans}), which transforms the time-dependent ancillary representation to the stationary one, can be written as
\begin{equation}\label{Vone}
\mathcal{V}(t)=S_1^\dagger\left[\theta(t)e^{-i\alpha(t)}\right]S_1\left[\theta(0)e^{-i\alpha(0)}\right].
\end{equation}
Using the Heisenberg equation~(\ref{vonNeu}) with the Hamiltonian~(\ref{HamNHsing}) and $\mathcal{J}_k(t)=\tilde{\mu}^\dagger(t)$, one can obtain the constraints for the driving intensity $\Omega(t)$ and eigenfrequency $\omega(t)$ as
\begin{equation}\label{CondiOne}
\begin{aligned}
&\Omega(t)=\frac{\dot{\theta}(t)-\gamma\sinh\theta(t)\cosh\theta(t)}{2\sin[\varphi+\alpha(t)]},\\
&\omega(t)=\left\{\dot{\alpha}(t)-2\Omega(t)\cos[\varphi+\alpha(t)]\right\}\coth2\theta(t).
\end{aligned}
\end{equation}
It is remarkable to find that Eq.~(\ref{CondiOne}) reduces to the constraints for the damping-free classical system that is allowed to explore the entire phase space when $\gamma=0$, $\alpha(t)=0$, and $\varphi=-\pi/2$ [see Eq.~(\ref{CondiClass})]. Subsequently, Eq.~(\ref{NonHermDynOri}) becomes
\begin{equation}\label{NonHermDynOriOne}
\hat{\mu}^\dagger(0)\rightarrow e^{if(t)}\hat{\mu}^\dagger(t)=e^{i[f_r(t)+f_i(t)]}\hat{\mu}^\dagger(t),
\end{equation}
where the time derivatives of the real and imaginary parts of the global phase are given by
\begin{equation}\label{global}
\begin{aligned}
\dot{f}_r(t)&=\omega(t)\sinh^2\theta(t)+\Omega(t)\cos[\varphi+\alpha(t)]\\
&\times\sinh2\theta(t)-\frac{1}{2}\dot{\alpha}(t)\sin2\theta(t),\\
\dot{f}_i(t)&=-i\frac{\gamma}{2}\sinh^2\theta(t).
\end{aligned}
\end{equation}
The last equation indicates that $f_i(t)$ is encoded with the dissipation process. To end up with a desired target state avoiding artificial normalization when $t=\tau$, one can design a two-stage gain or loss process:
\begin{equation}\label{imagphase}
\begin{aligned}
\gamma&=\lambda\left[\frac{\sinh2\theta(\tau)-\sinh2\theta(\tau_1)}{2\dot{\theta}(t)}-(\tau-\tau_1)\right],\quad t\in[0,\tau_1],\\
\gamma&=-\lambda\left[\frac{\sinh2\theta(\tau_1)-\sinh2\theta(0)}{2\dot{\theta}(t)}-\tau_1\right],\quad t\in[\tau_1,\tau]
\end{aligned}
\end{equation}
by imposing $f_i(\tau)=0$, where $\lambda>0$ scales the gain or loss rate and $\tau_1$ is an arbitrary intermediate moment. If we assume $\theta(0)=0$ and $|\psi(0)\rangle=|0\rangle$, then Eqs.~(\ref{Mone}) and (\ref{NonHermDynOriOne}) ensure that the control dynamics generates a squeezed vacuum state $|\psi_{\rm tar}(\tau)\rangle=S_1[\theta(\tau)\exp(-i\alpha(\tau))]|0\rangle$ as defined in Eq.~(\ref{SquezSing}).

\begin{figure}[htbp]
\centering
\includegraphics[width=0.9\linewidth]{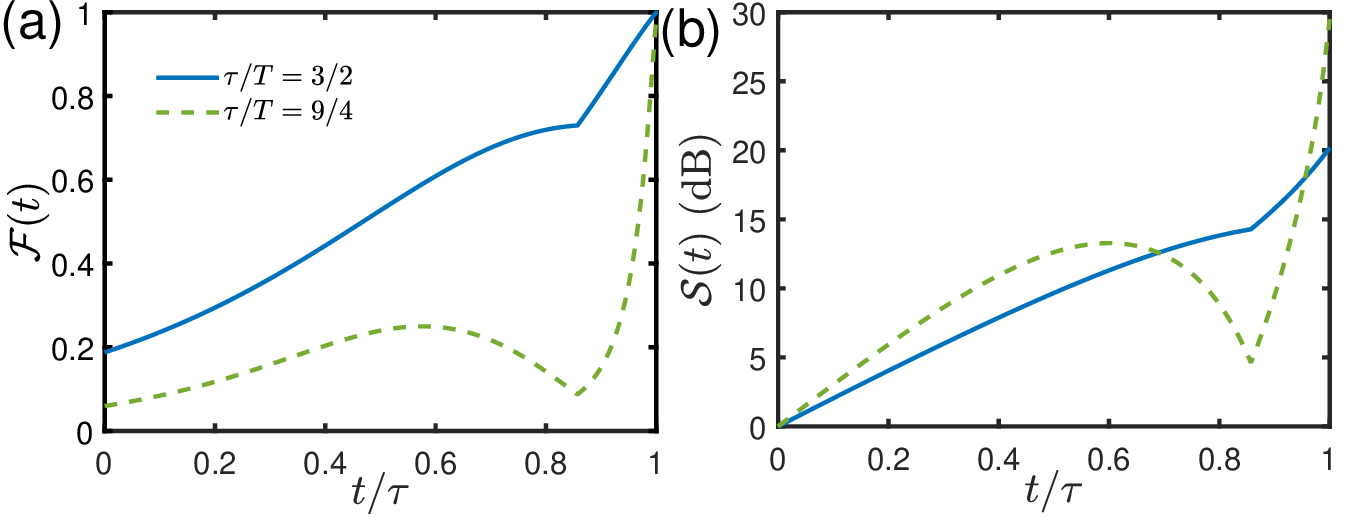}
\caption{Dynamics of (a) the fidelity with respect to the target single-mode squeezed state $|\psi_{\rm tar}(\tau)\rangle$ and (b) the squeezing level $\mathcal{S}(t)$. The Hilbert-space is truncated at $N=1400$. The driving intensity $\Omega(t)$ and the eigenfrequency $\omega(t)$ are set by Eq.~(\ref{CondiOne}), with $\varphi=\pi/2$, $\alpha(t)=0$, and $\theta(t)=\pi t/(2T)$ with a time constant $T\sim0.5\mu$s. The gain or loss rate $\gamma$ is defined in Eq.~(\ref{imagphase}) with $\tau_1=6\tau/7$. $\lambda=2$ for $\tau/T=3/2$ and $\lambda=0.2$ for $\tau/T=9/4$. $\gamma>0$ when $t\in[0,\tau_1]$ and $\gamma<0$ when $t\in[\tau_1,\tau]$. $|\gamma T|\sim0.5$ and $\Omega(t)\sim10-10^3\gamma$. All these parameters are experimentally practical in circuit-QED systems~\cite{Frattini2024Observation,Hajr2024High,Beaulieu2025Observation}.}\label{onesqueez}
\end{figure}

The performance of our protocol can be estimated by the fidelity $\mathcal{F}(t)=|\langle\psi(t)|\psi_{\rm tar}(\tau)\rangle|^2$ and the squeezing level $\mathcal{S}(t)=-10\log_{10}(2\Delta X^2)$ with $\Delta X^2\equiv\langle X^2\rangle-\langle X\rangle^2$ and $X=(\hat{a}_1+\hat{a}_1^\dagger)/\sqrt{2}$~\cite{Yuen1976Twophoton,Fabre2020Modes,Eriksson2024Universal,Cai2025Quantum} in Figs.~\ref{onesqueez}(a) and (b), respectively. The squeezing level and strength are correlated by $\mathcal{S}(t)=-10\log_{10}[\exp(-2r)]$ for an ideal squeezed state with $\phi=0$ in Eq.~(\ref{SquezSing}). The pure-state solution $|\psi(t)\rangle$ is numerically obtained by the time-dependent Schr\"odinger equation~(\ref{Sch}) in the ket space with the non-Hermitian Hamiltonian~(\ref{HamNHsing}). The consistency between non-Hermitian time-dependent Schr\"odinger equation and Lindblad master equation is numerically verified in Appendix~\ref{HeffDeriveSing}. The two-stage dynamics is illustrated in Fig.~\ref{onesqueez}, where the turning point is induced by the conversion of $\gamma$ in Eq.~(\ref{imagphase}) from positive (loss) to negative (gain). For both $\tau/T=3/2$ and $\tau/T=9/4$ in Fig.~\ref{onesqueez}(a), we obtain a unit fidelity at the target moment with a sufficient large truncated space. In Fig.~\ref{onesqueez}(b), the final squeezing level can be as high as $\mathcal{S}(\tau)\sim20.5$ dB for $\tau/T=3/2$ and $\mathcal{S}(\tau)\sim29.3$ dB for $\tau/T=9/4$, respectively, that allows a further enhancement for a longer coherent evolution.

\begin{figure}[htbp]
\centering
\includegraphics[width=0.9\linewidth]{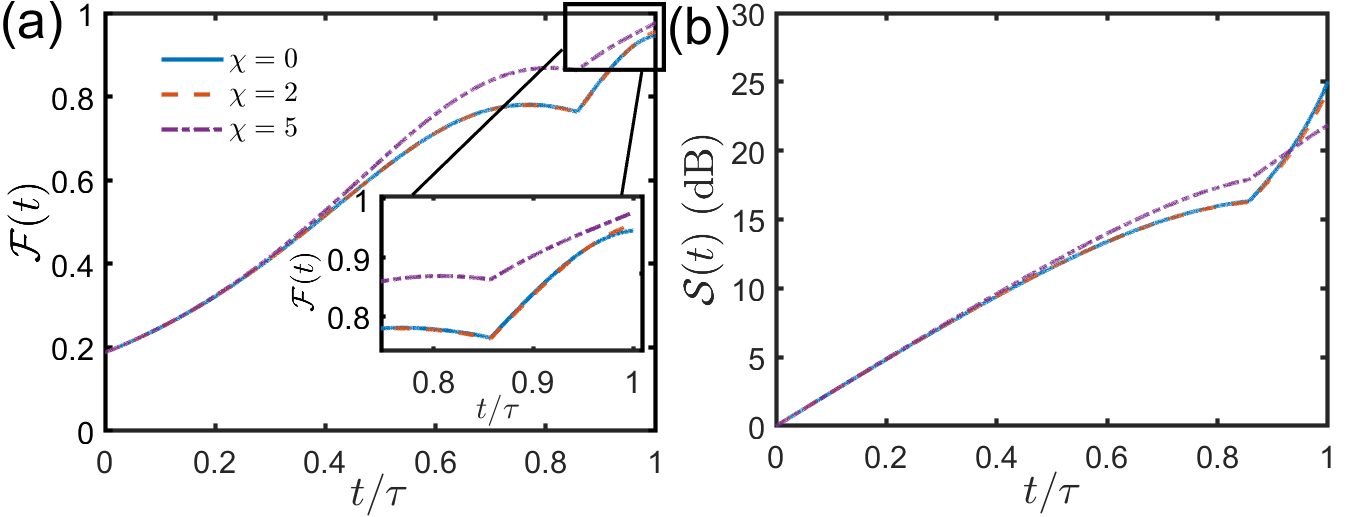}
\caption{Dynamics of (a) the fidelity with respect to the target single-mode squeezed state $|\psi_{\rm tar}(\tau)\rangle$ and (b) the squeezing level $\mathcal{S}(t)$ under the nonideal Hamiltonian in Eq.~(\ref{HamNon}) with a dimensionless deviation coefficient $\epsilon=0.2$. The Hilbert-space is truncated at $N=200$. $\varphi$ and $\alpha(t)$ are given by Eq.~(\ref{parameter}). The other parameters are the same as those for the case of $\tau/T=3/2$ in Fig.~\ref{onesqueez}.}\label{errsin}
\end{figure}

While our protocol for non-Hermitian Hamiltonian has considered the impact from external noise, its feasibility can be further discussed in the presence of parametric fluctuation. The fault tolerance of our universal passage in Eq.~(\ref{NonHermDynOriOne}) is enhanced by employing the dynamical error correction mechanism via modulating the time-dependent phase $\varphi(t)$ and the system eigenfrequency $\omega(t)$~\cite{Jin2025Dynamical}. Under the constraints in Eq.~(\ref{CondiOne}), the modulation can be simply realized by an extra setting
\begin{equation}\label{parameter}
\varphi(t)=-\alpha(t)+\frac{\pi}{2},\quad \alpha(t)=\frac{\chi}{100}\cosh[2\theta(t)],
\end{equation}
where $\chi$ is a coefficient for error correction. When $\chi=0$, the protocol restores to the original version.

Without loss of generality, one can consider that the non-Hermitian Hamiltonian~(\ref{HamNHsing}) fluctuated as
\begin{equation}\label{HamNon}
\hat{H}(t)\rightarrow\hat{H}(t)+\hat{H}_1(t),\quad \hat{H}_1(t)=\epsilon\left[\Omega(t)e^{i\varphi}\left(\hat{a}_1^\dagger\right)^2+{\rm H.c.}\right],
\end{equation}
where $\hat{H}_1(t)$ denotes the error Hamiltonian and the dimensionless coefficient $\epsilon$ indicates the deviation magnitude of the squeezing strength. Figure~\ref{errsin} demonstrates the effect of our dynamical error correction. In Fig.~\ref{errsin}(a), under a significant fluctuation in Eq.~(\ref{HamNon}) with $\epsilon=0.2$, the fidelity is $\mathcal{F}(\tau)=0.945$ for $\chi=0$, and increases to $\mathcal{F}(\tau)=0.956$ for $\chi=2$ and $\mathcal{F}(\tau)=0.977$ for $\chi=5$. Figure~\ref{errsin}(b) shows that the final squeezing level slightly decreases with increasing $\chi$. 

\section{Generation of double-mode squeezed state}\label{DoubleSquez}

In this section, we apply our dynamical-invariant-based protocol in Sec.~\ref{generalQuant} to the generation of a double-mode squeezed vacuum state, which is defined as~\cite{Gerry2023Introductory,Mamaev2025NonGaussian}
\begin{equation}\label{TwosquezDef}
\begin{aligned}
|\xi\rangle&\equiv S_2(\xi)|0\rangle_{a_1}\otimes|0\rangle_{a_2}=e^{\xi^*\hat{a}_1\hat{a}_2
-\xi\hat{a}_1^\dagger\hat{a}_2^\dagger}|0\rangle_{a_1}\otimes|0\rangle_{a_2}\\
&=\frac{1}{\cosh r}\sum_{n=0}^\infty\left(e^{-i\phi}\tanh r\right)^n|n\rangle_{a_1}\otimes|n\rangle_{a_2},
\end{aligned}
\end{equation}
with the double-mode squeezing operator $S_2(\xi)$ and the Fock state $|n\rangle_{a_k}$ of the $k$th mode. The non-Hermitian Hamiltonian for the double-mode bosonic system reads
\begin{equation}\label{HamTwo}
\hat{H}(t)=\sum_{k=1}^2\left[\omega_k(t)-i\frac{\gamma_k}{2}\right]\hat{a}_k^\dagger\hat{a}_k
+\left[g(t)e^{i\varphi}\hat{a}_1^\dagger\hat{a}_2^\dagger+{\rm H.c.}\right]
\end{equation}
with the eigenfrequency $\omega_k(t)$, the gain or loss rate $\gamma_k$ of the $k$th bosonic mode, the squeezing coupling strength $g(t)$ and the phase $\varphi$, which can also be derived from the Lindblad master equation (see Appendix~\ref{HeffDeriveTwo} for details). Distinct from the single-mode case, the two modes are required to be under loss and gain, respectively, which is available in certain systems, e.g., the cavity magnonic systems~\cite{Zhang2017Observation,Zhang2025Gain}. Due to the lack of the exchange interaction between the two modes, Eq.~(\ref{TransM}) can be simplified as $[\hat{\mu}_1(t),\hat{\mu}_2^\dagger(t)]^T=\mathcal{M}(t)(\hat{a}_1,\hat{a}_2^\dagger)^T$, where $\mathcal{M}(t)$ takes the same form as in Eq.~(\ref{Mone}). The completed version of $\mathcal{M}(t)$ for the double-mode system and a brief recipe about constructing the ancillary operators for a general $N$-mode system can be found in Appendix~\ref{recipe}. Then the unitary transformation $\mathcal{V}(t)$ that rotates from the time-dependent ancillary representation to the stationary one, i.e., $\hat{\mu}_1(t)\rightarrow\hat{\mu}_1(0)$ and $\hat{\mu}_2^\dagger(t)\rightarrow\hat{\mu}_2^\dagger(0)$, takes the similar form as Eq.~(\ref{Vone}) under the replacement of $S_1$ with $S_2$ in Eq.~(\ref{TwosquezDef}). Using the Heisenberg equation~(\ref{vonNeu}) with the Hamiltonian~(\ref{HamTwo}) and $\mathcal{J}_k(t)=\tilde{\mu}_2^\dagger(t)$, one can obtain the constraints for the squeezing coupling strength and eigenfrequency as
\begin{equation}\label{Condi}
\begin{aligned}
&g(t)=\frac{2\dot{\theta}(t)-(\gamma_1+\gamma_2)\sinh\theta(t)\cosh\theta(t)}{2\sin[\varphi+\alpha(t)]},\\
&\sum_{k=1}^2\omega_k(t)=2\left\{\dot{\alpha}(t)-g\cos[\varphi+\alpha(t)]\right\}\coth2\theta(t).
\end{aligned}
\end{equation}
Then according to Eq.~(\ref{NonHermDynOri}), the ancillary operator $\hat{\mu}_2^\dagger(t)$ evolves as
\begin{equation}\label{NonHermDynOritwo}
\hat{\mu}_2^\dagger(0)\rightarrow e^{if_2(t)}\hat{\mu}_2^\dagger(t)=e^{i[f_r(t)+f_i(t)]}\hat{\mu}_2^\dagger(t),
\end{equation}
where the time derivatives of the real and imaginary components of the global phase are
\begin{equation}\label{globaltwo}
\begin{aligned}
&\dot{f}_r(t)=\omega_1(t)\sinh^2\theta(t)+\omega_2(t)\cosh^2\theta(t)\\
&+g\cos[\varphi+\alpha(t)]\sinh2\theta(t)-\dot{\alpha}(t)\sinh2\theta(t),\\
&\dot{f}_i(t)=-\frac{i}{2}[\gamma_1\sinh^2\theta(t)+\gamma_2\cosh^2\theta(t)],
\end{aligned}
\end{equation}
respectively. The wavefunction probability can be conserved when $f_i(\tau)=0$ under certain conditions, e.g.,
\begin{equation}\label{condifi}
\gamma_{1,2}=\mp\lambda\left[\sinh2\theta(\tau)\pm2\dot{\theta}(t)\tau\right].
\end{equation}
They indicate that the gain or loss rate remains time-independent for a linear function of $\theta(t)$. For the initial conditions $\theta(0)=0$ and $|\psi(0)\rangle=|0\rangle_{a_1}\otimes|0\rangle_{a_2}$, the system dynamics described by Eq.~(\ref{NonHermDynOritwo}) generates a two-mode squeezed vacuum state $|\psi_{\rm tar}(\tau)\rangle=S_2[\theta(\tau)\exp{(-i\alpha(\tau))}]|\psi(0)\rangle$ as defined in Eq.~(\ref{TwosquezDef}).

\begin{figure}[htbp]
\centering
\includegraphics[width=0.9\linewidth]{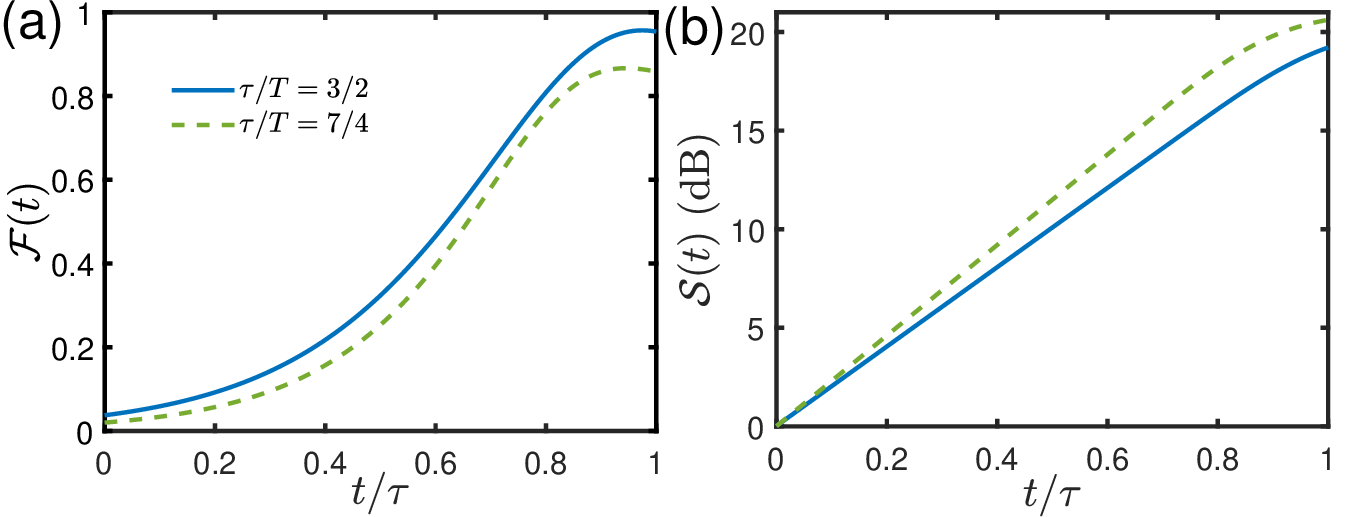}
\caption{Dynamics of (a) the fidelity with respect to the target two-mode squeezed vacuum state $|\psi_{\rm tar}(\tau)\rangle$ and (b) the relevant squeezing level $\mathcal{S}(t)$. The Hilbert-space of the two-mode system is truncated at $N=100\otimes100$. The squeezing coupling strength $g(t)$, and the eigenfrequencies $\omega_1(t)$ and $\omega_2(t)$ are chosen according to Eq.~(\ref{Condi}), where the gain or loss rates $\gamma_1$ and $\gamma_2$ are constrained by Eq.~(\ref{condifi}) with $\lambda=0.02$. $|\gamma_1T|\sim|\gamma_2T|\sim0.5$ and $g(t)\sim10\gamma_1$. The other parameters are the same as Fig.~\ref{onesqueez}. }\label{twosqueez}
\end{figure}

Figures~\ref{twosqueez}(a) and (b) demonstrate the dynamics of the fidelity about the target state and the relevant squeezing level, respectively. Double-mode squeezing level is defined as $\mathcal{S}(t)=-10\log_{10}(2\Delta X^2)$ with $X=(\hat{a}_1+\hat{a}_1^\dagger+\hat{a}_2+\hat{a}_2^\dagger)/2$. It is found that for target states $|\psi_{\rm tar}(\tau)\rangle$ with $\theta(\tau)=3\pi/4$ and $\theta(\tau)=7\pi/8$, the relevant squeezing levels are about $20$ dB and $25$ dB, respectively. In Fig.~\ref{twosqueez}(a), it is numerically confirmed that $\mathcal{F}(\tau)=0.954$ for $\tau/T=3/2$, and $\mathcal{F}(\tau)=0.860$ for $\tau/T=7/4$. And in Fig.~\ref{twosqueez}(b), the relevant squeezing levels can approach as high as $\mathcal{S}(\tau)\sim19.2$ dB for $\tau/T=3/2$ and $\mathcal{S}(\tau)\sim20.6$ dB for $\tau/T=7/4$, respectively.

In comparison to the single-mode result in Fig.~\ref{onesqueez}, the reduced fidelity in the double-mode system is mainly due to the constrained memory size and computational capability of our laptop [an Intel Core i5-10400F processor with $2.90$ gigahertz in frequency and $32$ gigabyte in memory] rather than the theory itself. The analytical derivation is valid in the whole Hilbert space while the numerical simulations can only be performed in a truncated space. For the single-mode case, a size of about $N=1400$ Fock bases is sufficient to achieve a unit fidelity. While for the double-mode case, we are reasonable to expect a larger fidelity and a higher squeezing level in a greater size of truncated space.

\section{Conclusion}\label{Conclusion}

We establish a unified framework based on dynamical invariants in control of both classical and quantum continuous-variable systems under time-dependent Hamiltonian. It is embedded in the differential manifolds of the time-dependent ancillary canonical variables or operators. Our universal quantum control appears as a natural consequence of the Hamilton-Jacobi equation for classical systems. The time-dependent ancillary variables, which is obtained by the generating function for canonical transformation, serve as dynamical invariants for controlling the expansion and transport of the system trajectories in the phase space. As the dynamical invariants for quantum systems, the time-dependent ancillary operators are obtained by the same symplectic transformation as in classical systems. Then the unitary transformation from the time-dependent ancillary operators to the stationary ones induces a general gauge potential that encodes the information of differential structure. The second quantization offers a deeper insight by mapping the Liouville equation to the Heisenberg equation, which serves as a sufficient condition to construct both nonadiabatic passages toward the target modes and exact solutions to the time-dependent Schr\"odinger equation in either Hermitian or non-Hermitian quantum mechanics. Our work therefore constitutes the groundwork for the whole framework of universal quantum control by connecting analytical mechanics and quantum control.

As a nontrivial application, we apply our theory to single- and double-mode bosonic systems governed by the non-Hermitian squeezing Hamiltonian derived from the Lindblad master equation rather than phenomenologically assumed. The universal passages are demonstrated by dynamical invariants satisfying the Heisenberg equation with the non-Hermitian Hamiltonian, thereby enabling arbitrary target states in the presence of noise. The generated single- and double-mode squeezed states achieve ultrahigh squeezing levels of $29.3$ dB and $20.6$ dB, respectively. Moreover, the fault-tolerance of the universal passages against parametric fluctuations can be enhanced by modulating the time-dependent phase and bosonic-mode eigenfrequency.

\section*{Acknowledgments}

We acknowledge grant support from the National Natural Science Foundation of China (Grants No. U25A20199 and No. 92265202), the National Key Research and Development Program of China (Grant No. 2022YFA1405200), and the ``Pioneer'' and ``Leading Goose'' R\&D of Zhejiang Province (Grant No. 2025C01028).

\bibliographystyle{apsrevlong}
\bibliography{ref}


\appendix
\onecolumngrid


\setcounter{section}{0}
\renewcommand{\thesection}{\Alph{section}}
\setcounter{subsection}{0}
\makeatletter
\renewcommand{\thesubsection}{\Alph{section}\arabic{subsection}}
\renewcommand{\p@subsection}{} 
\makeatother
\setcounter{equation}{0}
\renewcommand{\theequation}{\thesection\arabic{equation}}

\makeatletter
\@addtoreset{equation}{section} 
\makeatother

\section{Canonical transformation and Hamilton-Jacobi equation}\label{UCCone}

In this section, we use canonical transformation and the Hamilton-Jacobi equation to construct the dynamical invariants of single- and double-resonator classical systems as the desired time-dependent ancillary canonical variables, enabling the systems to reach arbitrary locations in phase space via nonadiabatic trajectories.

\subsection{Single-mode classical systems}

Consider a general time-dependent Hamiltonian~\cite{Yeon1993Exact,Harari2011Propagator,Jarzynski2013Generating,
Deffner2014Classical,Harari2015Quantum,Anzaldo2017Quadratic} for the single-mode classical systems:
\begin{equation}\label{Hamclass}
H(q,p,t)=\frac{\lambda(t)p^2}{2m}+\frac{1}{2}\chi(t)m\omega^2q^2+2\Omega(t)\cos(2\omega t)qp,
\end{equation}
where $q$ and $p$ denote the canonical coordinate and momentum, respectively. The time-dependent parameters $\lambda(t)$ and $\chi(t)$ control the kinetic and potential energies, respectively; $\Omega(t)$ determines the coupling strength between $q$ and $p$. The dynamics of classical system governed by Eq.~(\ref{Hamclass}) can be mapped to the control of a lattice of particles and a gas of individual particles~\cite{Jarzynski2013Generating}. For example, consider a lattice of particles distributed at equal intervals within a box of a length $L$, under a potential whose magnitude is finite inside the box and infinite outside. These particles are subject to a velocity-dependent external force, represented by the last term $\sim qp$ in Eq.~(\ref{Hamclass}). The lattice can be uniformly stretched or contracted in analog to the accordion as the box length is varied. Similarly, controlled by Eq.~(\ref{Hamclass}), a gas of independent particles sampled at equally spaced intervals in a box can remain uniformly distributed even when the box length varies with time in an arbitrary way.

The classical Hamiltonian in Eq.~(\ref{Hamclass}) can be expressed by a more compact form, under the time-independent canonical transformation $(q,p)\rightarrow(a,a^*)$ with the new canonical variables as~\cite{STROCCHI1966Complex}
\begin{equation}\label{CanTransa}
a=\frac{1}{\sqrt{2}}\left(\sqrt{m\omega}q+\frac{i}{\sqrt{m\omega}}p\right)
\end{equation}
and its complex conjugate $a^*$, and they satisfy the Poisson bracket relations as
\begin{equation}\label{Poissona}
\left\{a,a^*\right\}_{q,p}=-i,\quad \left\{a,a\right\}_{q,p}=\left\{a^*,a^*\right\}_{q,p}=0.
\end{equation}
Under the conditions of $\lambda(t)=1-2\Omega(t)\sin(2\omega t)/\omega$ and $\chi(t)=1+2\Omega(t)\sin(2\omega t)/\omega$, we find that the Hamiltonian $H(q,p,t)$ in Eq.~(\ref{Hamclass}) can be expressed in a similar form to the quantum squeezing Hamiltonian
\begin{equation}\label{Hama}
H(a,a^*,t)=\omega a^*a+i\Omega(t)\left({a^*}^2e^{-i2\omega t}-a^2e^{i2\omega t}\right),
\end{equation}
which allows the mechanical parametric amplification and the thermomechanical noise squeezing~\cite{Rugar1991Mechanical}. Due to Eq.~(\ref{Poissona}), the Hamilton's canonical equations of $a$ and $a^*$ can be derived as~\cite{STROCCHI1966Complex}
\begin{equation}\label{HamSingle}
i\frac{da}{dt}=\frac{\partial H(a,a^*,t)}{\partial a^*},\quad i\frac{da^*}{dt}=-\frac{\partial H(a,a^*,t)}{\partial a}.
\end{equation}

We adopt an alternative perspective to obtain the system dynamics rather than directly solving Eq.~(\ref{HamSingle}). To have a pure ``squeezing'' Hamiltonian, we introduce the first time-dependent canonical transformation as
\begin{equation}\label{TransSinb}
b=-ae^{i\omega t},\quad b^*=-a^*e^{-i\omega t},
\end{equation}
which satisfies the Poisson bracket $\{b,b^*\}_{q,p}=-i$. The Hamilton's canonical equations for $b$ and $b^*$ are
\begin{equation}\label{Canob}
\frac{db}{dt}=-i\frac{\partial H_{\rm rot}(b,b^*,t)}{\partial b^*},\quad \frac{db^*}{dt}=i\frac{\partial H_{\rm rot}(b,b^*,t)}{\partial b}
\end{equation}
with the transformed Hamiltonian
\begin{equation}\label{HamConstr}
H_{\rm rot}(b,b^*,t)=H(a,a^*,t)+\frac{\partial F_3^{(1)}(a,b^*,t)}{\partial t},
\end{equation}
where the type-III generating function $F_3^{(1)}(a,b^*,t)$ is constrained by
\begin{equation}\label{F2constrain}
a^*=-i\frac{\partial F_3^{(1)}(a,b^*,t)}{\partial a},\quad b=-i\frac{\partial F_3^{(1)}(a,b^*,t)}{\partial b^*}.
\end{equation}
Equations~(\ref{HamConstr}) and (\ref{F2constrain}) follow the Hamilton's variational principle~\cite{Arnol2013mathematical,Landau2013course,STROCCHI1966Complex}
\begin{equation}\label{Hamprinc}
\delta\int_0^t\left[ia\dot{a}^*-H(a,a^*,t)\right]dt=\delta\int_0^t\left[ib\dot{b}^*-H_{\rm rot}(b,b^*,t)+\frac{dF_3^{(1)}(a,b^*,t)}{dt}\right]dt=0,
\end{equation}
where the total time derivative of the generating function is $dF_3^{(1)}(a,b^*,t)/dt=\partial_aF_3^{(1)}\dot{a}+\partial_{b^*}F_3^{(1)}\dot{b}^*+\partial_tF_3^{(1)}$.

Using the quadratic ansatz $F_3^{(1)}(a,b^*,t)=A_1a^2+B_1ab^*+C_1{b^*}^2$ with undetermined coefficients $A_1$, $B_1$, and $C_1$ and Eqs.~(\ref{TransSinb}) and (\ref{F2constrain}), we have
\begin{equation}\label{GneraFunb}
F_3^{(1)}(a,b^*,t)=-iab^*e^{i\omega t}.
\end{equation}
Then using Eqs.~(\ref{Hama}), (\ref{TransSinb}), and (\ref{GneraFunb}), the first-rotated Hamiltonian in Eq.~(\ref{HamConstr}) can be expressed as
\begin{equation}\label{HamClab}
\begin{aligned}
H_{\rm rot}(b,b^*,t)&=H(a,a^*,t)+\frac{\partial F_3^{(1)}(a,b^*,t)}{\partial t}=\omega a^*a+i\Omega(t)\left({a^*}^2e^{-i2\omega t}-a^2e^{i2\omega t}\right)+\omega ab^*e^{i\omega t}\\
&=\omega bb^*+i\Omega(t)({b^*}^2-b^2)-\omega bb^*=i\Omega(t)\left({b^*}^2-b^2\right).
\end{aligned}
\end{equation}

In preceding, we consider the second time-dependent canonical transformation as
\begin{equation}\label{TranSingle}
[u(b,b^*,t),u^*(b,b^*,t)]^T=\mathcal{M}(t)(b,b^*)^T,
\end{equation}
where the time-dependent $2\times2$ symplectic matrix $\mathcal{M}(t)$ reads,
\begin{equation}\label{TransMsingle}
\mathcal{M}(t)=\begin{pmatrix}\cosh\theta(t)&-\sinh\theta(t)\\-\sinh\theta(t)&\cosh\theta(t)\end{pmatrix},
\end{equation}
One can verify that the ancillary canonical variables satisfy $\{u,u^*\}_{q,p}=-i$. Similar to Eq.~(\ref{F2constrain}), the second-rotated generating function for Eqs.~(\ref{TranSingle}) and (\ref{TransMsingle}) is constrained by $b^*=-i\partial F_3^{(2)}/\partial b$ and $u=-i\partial F_3^{(2)}/\partial u^*$. Assuming a quadratic form for the generating function $F_3^{(2)}(b,u^*,t)=A_2b^2+B_2bu^*+C_2{u^*}^2$ with the coefficients $A_2$, $B_2$, and $C_2$ yet to be determined, one can find
\begin{equation}\label{F2form}
F_3^{(2)}(b,u^*,t)=\frac{i\tanh\theta(t)}{2}(b^2-{u^*}^2)+\frac{i}{\cosh\theta(t)}bu^*.
\end{equation}
The second-transformed Hamiltonian can be obtained as
\begin{equation}\label{HamKSingle}
\begin{aligned}
K(u,u^*,t)&=H_{\rm rot}(b,b^*,t)+\frac{\partial F_3^{(2)}(b,u^*,t)}{\partial t}=i\Omega(t)({b^*}^2-b^2)+i\dot{\theta}(t)
\left[\frac{1}{2\cosh^2\theta(t)}(b^2-{u^*}^2)-\frac{\sinh\theta(t)}{\cosh^2\theta(t)}bu^*\right]\\
&=i\Omega(t)\left({u^*}^2-u^2\right)-i\frac{\dot{\theta}(t)}{2}\left({u^*}^2-u^2\right),
\end{aligned}
\end{equation}
where the last equality is attained by Eq.~(\ref{TranSingle}). When Eq.~(\ref{HamKSingle}) is cast into the Hamilton-Jacobi equation, i.e., $K(u,u^*,t)=0$, one can obtain the constraint for $\Omega(t)$ as
\begin{equation}\label{CondiClass}
\Omega(t)=\frac{\dot{\theta}(t)}{2}.
\end{equation}
Remarkably, under $\gamma=0$, $\alpha(t)=0$, and $\varphi=-\pi/2$, the two-photon driving intensity in Eq.~(\ref{CondiOne}) used for generating the squeezed state by a quantum Hermitian system follows the classical condition in Eq.~(\ref{CondiClass}). Since the ancillary canonical variables $u$ and $u^*$ are dynamical invariants, the system trajectory in the phase space $(a,a^*)$ can be directly obtained by Eqs.~(\ref{TransSinb}) and (\ref{TranSingle}).

\subsection{Double-mode classical systems}

Consider a two-mode classical system described by the time-dependent Hamiltonian~\cite{Briggs2012Coherent,Briggs2013Quantum,Liu2024NonAbelian} as
\begin{equation}\label{Hamclasstwo}
\begin{aligned}
&H(\vec{q},\vec{p},t)=\frac{p_1^2+p_2^2}{2m}+\frac{1}{2}m\omega^2\left(q_1^2+q_2^2\right)-\Omega(t)\cos(2\omega t)(q_1p_2+p_1q_2)+\Omega(t)\sin(2\omega t)\left(m\omega q_1q_2-\frac{1}{m\omega}p_1p_2\right),
\end{aligned}
\end{equation}
where $\vec{q}=(q_1,q_2)$ and $\vec{p}=(p_1,p_2)$, with $k=1,2$, denote the canonical coordinates and momenta, respectively. $\Omega(t)$ is the time-dependent coupling strength between the coordinates and momenta of the two harmonic oscillators. The dynamics of classical systems governed by Eq.~(\ref{Hamclasstwo}) can be mapped to the solutions of two-level quantum systems, enabling the generation of eigenstates~\cite{Briggs2012Coherent} and the implementation of quantum gates~\cite{Briggs2013Quantum}.

Under the time-independent canonical transformation in Eq.~(\ref{CanoTrans}) with $N=2$, $\omega_1=\omega_2=\omega$, and $m_1=m_2=m$ or Eq.~(\ref{CanTransa}), the Hamiltonian in Eq.~(\ref{Hamclasstwo}) can be written in a compact form analogous to a two-mode quantum squeezing Hamiltonian as
\begin{equation}\label{HamaTwo}
H(\vec{a},\vec{a}^*,t)=\omega(a_1^*a_1+a_2^*a_2)+i\Omega(t)\left(a_1^*a_2^*e^{-i2\omega t}-a_1a_2e^{i2\omega t}\right),
\end{equation}
with $\vec{a}=(a_1,a_2)$ and $\vec{a}^*=(a_1^*,a_2^*)$. The Hamilton's canonical equations of $a_k$ and $a_k^*$ with $k=1,2$ can then be obtained as
\begin{equation}\label{HamTwoClas}
i\frac{da_k}{dt}=\frac{\partial H(\vec{a},\vec{a}^*,t)}{\partial a_k^*},\quad i\frac{da_k^*}{dt}=-\frac{\partial H(\vec{a},\vec{a}^*,t)}{\partial a_k}.
\end{equation}
Regarding a canonical transformation similar to Eq.~(\ref{TransSinb}), i.e.,
\begin{equation}\label{TransTwob}
b_k=-a_ke^{i\omega t},\quad b_k^*=-a_k^*e^{-i\omega t},\quad k=1,2,
\end{equation}
the type-III generating function is constrained by $a_k^*=-i\partial F_3^{(1)}/\partial a_k$ and $b_k=-i\partial F_3^{(1)}/\partial b_k^*$. Assuming a quadratic ansatz $F_3^{(1)}(\vec{a},\vec{b}^*,t)=A_1a_1a_2+B_1(a_1b_1^*+a_2b_2^*)+C_1{b_1^*b_2^*}$, the generating function takes the form of
\begin{equation}\label{GneraFunbtwo}
F_3^{(1)}(\vec{a},\vec{b}^*,t)=-i(a_1b_1^*+a_2b_2^*)e^{i\omega t}.
\end{equation}
Using Eqs.~(\ref{HamaTwo}) and (\ref{GneraFunbtwo}), the first-rotated Hamiltonian can be obtained as
\begin{equation}\label{HamClabTwo}
\begin{aligned}
H_{\rm rot}(\vec{b},\vec{b}^*,t)&=H(\vec{a},\vec{a}^*,t)+\frac{\partial F_3^{(1)}(\vec{a},\vec{b}^*,t)}{\partial t}=\omega(a_1^*a_1+a_2^*a_2)+i\Omega(t)(a_1^*a_2^*e^{-i2\omega t}-a_1a_2e^{i2\omega t})\\
&+\omega(a_1b_1^*+a_2b_2^*)e^{i\omega t}=i\Omega(t)(b_1^*b_2^*-b_1b_2),
\end{aligned}
\end{equation}
where the last step is derived through the inverse transformation of Eq.~(\ref{TransTwob}).

Similar to Eqs.~(\ref{TransR}) and (\ref{TranSingle}), we consider the time-dependent canonical transformation as
\begin{equation}\label{TransTwo}
[u_1(b_1,b_2^*,t),u_2^*(b_1,b_2^*,t)]^T=\mathcal{M}(t)(b_1,b_2^*)^T,
\end{equation}
with the time-dependent symplectic matrix $\mathcal{M}(t)$ defined in Eq.~(\ref{TransMsingle}). Under the constraints of the type-III generating function $F_3^{(2)}(\vec{b},\vec{u}^*,t)$, i.e., $b_k^*=-i\partial F_3^{(2)}/\partial b_k$ and $u_k=-i\partial F_3^{(2)}/\partial u_k^*$, and the assumption of quadratic ansatz $F_3^{(2)}(\vec{b},\vec{u}^*,t)=A_2b_1b_2+B_2(b_1u_1^*+b_2u_2^*)+C_2{u_1^*u_2^*}$, we have
\begin{equation}\label{F2formTwo}
F_3^{(2)}(\vec{b},\vec{u}^*,t)=i\tanh\theta(t)(b_1b_2-u_1^*u_2^*)+\frac{i}{\cosh\theta(t)}(b_1u_1^*+b_2u_2^*).
\end{equation}
Using Eqs.~(\ref{TransTwo}) and (\ref{F2formTwo}), $H_{\rm rot}(\vec{b},\vec{b}^*,t)$ in Eq.~(\ref{HamClabTwo}) can be further transformed to
\begin{equation}\label{HamKTwo}
\begin{aligned}
&K(\vec{u},\vec{u}^*,t)=H_{\rm rot}(\vec{b},\vec{b}^*,t)+\frac{\partial F_3^{(2)}(\vec{b},\vec{u}^*,t)}{\partial t}=i\Omega(t)(b_1^*b_2^*-b_1b_2)+i\frac{\dot{\theta}(t)}{\cosh^2\theta(t)}(b_1b_2-u_1^*u_2^*)\\
&-i\frac{\dot{\theta}(t)\sinh\theta(t)}{\cosh^2\theta(t)}(b_1u_1^*+b_2u_2^*)
=i\Omega(t)(u_1^*u_2^*-u_1u_2)+i\dot{\theta}(t)(u_1u_2-u_1^*u_2^*).
\end{aligned}
\end{equation}
On substituting Eq.~(\ref{HamKTwo}) into the Hamilton-Jacobi equation, i.e., $K(u,u^*,t)=0$, the constraint for $\Omega(t)$ can be obtained as
\begin{equation}\label{CondiTwo}
\Omega(t)=\dot{\theta}(t).
\end{equation}
Interestingly, the classical constraint in Eq.~(\ref{CondiTwo}) coincides with that for the coupling strength in Eq.~(\ref{Condi}) about the double-mode squeezing interaction in the quantum system, under the conditions of $\gamma_1=0$, $\gamma_2=0$, $\alpha=0$, and $\varphi=-\pi/2$. Similar to the case of the single harmonic oscillator, the trajectories of two coupled harmonic oscillators in the phase space $(a_1,a_2^*)$ or $(a_1^*,a_2)$ can be fully determined by Eqs.~(\ref{TransTwob}) and (\ref{TransTwo}).

\section{Derivation of Liouville equation~(\ref{vonNeuClas})}\label{DervonNeuClas}

This section discusses the general $N$-mode classical systems, which are described by the time-dependent ancillary canonical variables $u_k(\vec{a},\vec{a}^*,t)$ and $u_k^*(\vec{a},\vec{a}^*,t)$, $1\leq k\leq N$. Using the Hamilton-Jacobi equations that generalizes the single- and double-mode cases in \red{Appendix~\ref{UCCone}}, one can obtain the Liouville equation~(\ref{vonNeuClas}).

By accounting for the implicit time-dependence of $\vec{a}$ and $\vec{a}^*$ in $u_k(\vec{a},\vec{a}^*,t)$ and $u_k^*(\vec{a},\vec{a}^*,t)$ and recalling the Hamilton's canonical equations of $u_k$ and $u_k^*$ in Eq.~(\ref{HamKEq}) as
\begin{equation}\label{HamKEqApp}
\frac{du_k}{dt}=-i\frac{\partial K(\vec{u},\vec{u}^*,t)}{\partial u_k^*}=0,\quad \frac{du_k^*}{dt}=i\frac{\partial K(\vec{u},\vec{u}^*,t)}{\partial u_k}=0,
\end{equation}
one can find that
\begin{equation}\label{QPexplict}
\begin{aligned}
&\frac{du_k(\vec{a},\vec{a}^*,t)}{dt}=\sum_{k=1}^N\left[\frac{\partial u_k(\vec{a},\vec{a}^*,t)}{\partial a_k}\frac{da_k}{dt}+\frac{\partial u_k(\vec{a},\vec{a}^*,t)}{\partial a_k^*}\frac{da_k^*}{dt}\right]+\frac{\partial u_k(\vec{a},\vec{a}^*,t)}{\partial t}=0,\\
&\frac{du_k^*(\vec{a},\vec{a}^*,t)}{dt}=\sum_{k=1}^N\left[\frac{\partial u_k^*(\vec{a},\vec{a}^*,t)}{\partial a_k}\frac{da_k}{dt}+\frac{\partial u_k^*(\vec{a},\vec{a}^*,t)}{\partial a_k^*}\frac{da_k^*}{dt}\right]+\frac{\partial u_k^*(\vec{a},\vec{a}^*,t)}{\partial t}=0.
\end{aligned}
\end{equation}
Using the Hamilton's canonical equation~(\ref{HamEq}) and the Poisson bracket  $\{A,B\}_{\vec{a},\vec{a}^*}=\sum_{k=1}^N(\partial_{a_k}A\partial_{\partial a_k^*}B-\partial_{a_k^*}A\partial_{a_k}B)$ with $A$ and $B$ arbitrary variables, Eq.~(\ref{QPexplict}) can be transformed as
\begin{equation}\label{Qpartial}
\begin{aligned}
\frac{\partial u_k(\vec{a},\vec{a}^*,t)}{\partial t}&=-\sum_{k=1}^N\Big[\frac{\partial u_k(\vec{a},\vec{a}^*,t)}{\partial a_k}\frac{da_k}{dt}+\frac{\partial u_k(\vec{a},\vec{a}^*,t)}{\partial a_k^*}\frac{da_k^*}{dt}\Big]=i\sum_{k=1}^N\Big[\frac{\partial u_k(\vec{a},\vec{a}^*,t)}{\partial a_k}\frac{\partial H(\vec{a},\vec{a}^*,t)}{\partial a_k^*}\\
&-\frac{\partial u_k(\vec{a},\vec{a}^*,t)}{\partial a_k^*}\frac{\partial H(\vec{a},\vec{a},t)}{\partial a_k}\Big]=-i\left\{H(\vec{a},\vec{a}^*,t),u_k(\vec{a},\vec{a}^*,t)\right\}_{\vec{a},\vec{a}^*},
\end{aligned}
\end{equation}
and
\begin{equation}\label{Ppartial}
\begin{aligned}
\frac{\partial u_k^*(\vec{a},\vec{a}^*,t)}{\partial t}&=-\sum_{k=1}^N\Big[\frac{\partial u_k^*(\vec{a},\vec{a}^*,t)}{\partial a_k}\frac{da_k}{dt}+\frac{\partial u_k^*(\vec{a},\vec{a}^*,t)}{\partial a_k^*}\frac{da_k^*}{dt}\Big]=i\sum_{k=1}^N\Big[\frac{\partial u_k^*(\vec{a},\vec{a}^*,t)}{\partial a_k}\frac{\partial H(\vec{a},\vec{a}^*,t)}{\partial a_k^*}\\
&-\frac{\partial u_k^*(\vec{a},\vec{a}^*,t)}{\partial a_k^*}\frac{\partial H(\vec{a},\vec{a},t)}{\partial a_k}\Big]=-i\left\{H(\vec{a},\vec{a}^*,t),u_k^*(\vec{a},\vec{a}^*,t)\right\}_{\vec{a},\vec{a}^*}.
\end{aligned}
\end{equation}
Equations~(\ref{Qpartial}) and (\ref{Ppartial}) take the same form as the Liouville equation~(\ref{vonNeuClas}) with $J_k=u_k$ and $u_k^*$, respectively.

\section{Hamilton-Jacobi equation for classical control}\label{HJscale}

This section shows that the Hamilton-Jacobi equation, which leads to the Liouville equation~(\ref{vonNeuClas}), can be used to construct the ancillary Hamiltonian for the control of classical scale-invariant systems~\cite{Jarzynski2013Generating,Deffner2014Classical,Jarzynski2017Fast}. It encompasses the classical shortcut-to-adiabaticity proposals that guide all the system trajectories in phase space with a given initial action to end up with the same value of action~\cite{Jarzynski2013Generating,Deffner2014Classical,Jarzynski2017Fast}. The controllability of these trajectories can be mapped to the control of the transport, expansion, and compression of a quantum harmonic oscillator in a time-dependent trap~\cite{Jarzynski2013Generating,Deffner2014Classical,Jarzynski2017Fast}.

The classical scale-invariant system is generally governed by the following time-dependent Hamiltonian~\cite{Deffner2014Classical,Jarzynski2017Fast},
\begin{equation}\label{HamScale}
H(q,p,t)=\frac{p^2}{2m}+U(q,t).
\end{equation}
Under a time-dependent canonical transformation $(q,p)\rightarrow[Q(q,p,t),P(q,p,t)]$ based on a special type-II generating function $F_2(q,t)$ that explicitly depends only on $q$ and $t$, the transformed Hamiltonian can be written as
\begin{equation}\label{HamScaleTrans}
K(Q,P,t)=H(q,p,t)+\frac{\partial F_2(q,t)}{\partial t}.
\end{equation}
Using the condition of the Hamilton-Jacobi equation, i.e., $K(Q,P,t)=0$, we have
\begin{equation}\label{CondiHam}
\partial_tF_2(q,t)+\frac{[\partial_qF_2(q,t)]^2}{2m}+U(q,t)=0,
\end{equation}
which is subject to the generating function $p=\partial_qF_2(q,t)$. The partial derivative of Eq.~(\ref{CondiHam}) with respect to $q$ yields
\begin{equation}\label{CondiHamPart}
\partial_qU(q,t)=-\partial_q\partial_tF_2(q,t)-\frac{1}{m}\partial_qF_2(q,t)
\partial_q^2F_2(q,t)=-\partial_t\left[mv(q,t)\right]-mv(q,t)\partial_qv(q,t)=-ma(q,t),
\end{equation}
in which we use the definitions of the velocity and acceleration fields~\cite{Jarzynski2017Fast}
\begin{equation}\label{VA}
v(q,t)\equiv\frac{\partial_qF_2(q,t)}{m},\quad a(q,t)\equiv \frac{dv(q,t)}{dt}=\partial_q v\frac{dq}{dt}+\partial_tv.
\end{equation}

Substitute the quadratic ansatz $F_2(q,t)=A(t)[q-\beta(t)]^2+B(t)[q-\beta(t)]+C(t)$ with the time-dependent coefficients $A(t)$, $B(t)$, and $C(t)$, and the scale-invariant potential $U(q,t)=D(t)[(q-\beta(t))/\eta(t)]^2+E(t)q$ with the coefficients $D(t)$ and $E(t)$ into Eq.~(\ref{CondiHamPart}), the generating function $F_2(q,t)$ is found to be
\begin{equation}\label{F2scale}
F_2(q,t)=m\frac{\dot{\eta}(t)}{2\eta(t)}[q-\beta(t)]^2+m\dot{\beta}(t)[q-\beta(t)]+\int_0^t\frac{m}{2}\dot{\beta}^2(s)ds.
\end{equation}
Then the velocity and acceleration fields in Eq.~(\ref{VA}) read
\begin{equation}\label{VAsov}
v(q,t)=\frac{\dot{\eta}(t)}{\eta(t)}[q-\beta(t)]+\dot{\beta}(t),\quad a(q,t)=\frac{\ddot{\eta}(t)}{\eta(t)}[q-\beta(t)]+\ddot{\beta}(t),
\end{equation}
respectively. According to Eqs.~(\ref{CondiHamPart}) and (\ref{VAsov}), the potential can be expressed as
\begin{equation}\label{Uscale}
U(q,t)=-\int_0^qma(q',t)dq'=-\frac{m}{2}\frac{\ddot{\eta}(t)}{\eta(t)}[q-\beta(t)]^2-m\ddot{\beta}(t)q.
\end{equation}
Equation~(\ref{Uscale}) takes the same form as the fast-forward potential in Eq.~(24) of the classical shortcut-to-adiabaticity proposal obtained by Jarzynski et al.~\cite{Jarzynski2017Fast}.

\section{Universal quantum control with time-dependent Hermitian Hamiltonian}\label{SufNecPartDiag}

\subsection{Quadratic Hamiltonian}\label{quadratic}

This subsection shows that when both $\tilde{\mu}_k(t)\equiv e^{if_k(t)}\hat{\mu}_k(t)$ and $\tilde{\mu}_k^\dagger(t)\equiv e^{-if_k(t)}\hat{\mu}_k^\dagger(t)$ in Eq.~(\ref{Quanti}) satisfy the Heisenberg equation~(\ref{vonNeu}) with the time-dependent Hamiltonian $\hat{H}(t)$ in Eq.~(\ref{Ham}) or (\ref{HamMu}), they suffice to demonstrate the controllable dynamics in Eq.~(\ref{HermDynOri}) for quantum Hermitian systems.

For simplicity, we consider the general system of $N$ coupled bosonic modes only with squeezing interactions. The time-dependent Hamiltonian (\ref{HamMu}) in the ancillary representation is then given by
\begin{equation}\label{HamMuAppend}
\hat{H}(t)=\begin{pmatrix}\vec{\mu}^\dagger&\vec{\mu}\end{pmatrix}H^\mu(t)
\begin{pmatrix}\vec{\mu}^T\\(\vec{\mu}^\dagger)^T\end{pmatrix},
\end{equation}
where the $2N\times2N$ coefficient matrix $H^\mu(t)$ can be divided into four $N\times N$ blocks as
\begin{eqnarray}
\begin{aligned}
H^\mu(t)&=\left(\begin{array}{c|c}H^\mu_d(t)&H^\mu_s(t)\\
    \hline
    (H^\mu_s(t))^\dagger&\bf{0}\end{array}\right)\\
    &=\left(\begin{array}{ccccc|ccccc}
	H^\mu_{1,1}(t) & 0 & \cdots & 0 & 0 & 0 & H^\mu_{1,N+2}(t) & \cdots & H^\mu_{1,2N-1}(t) & H^\mu_{1,2N}(t)\\
	0 & H^\mu_{2,2}(t) & \cdots & 0 & 0 &0 & 0 & \cdots & H^\mu_{2,2N-1}(t) & H^\mu_{2,2N}(t)\\
	\vdots & \vdots & \ddots & \vdots & \vdots & \vdots & \vdots & \ddots &\vdots & \vdots\\
	0 & 0 & \cdots & H^\mu_{N-1,N-1}(t) & 0 & 0 & 0 & \cdots & 0 & H^\mu_{N-1,2N}(t)\\
	0 & 0 & \cdots & 0 & H^\mu_{N,N}(t) & 0 & 0 & \cdots & 0 & 0\\
	\hline
	0 & 0 & \cdots & 0 & 0 & 0 & 0 & \cdots & 0 & 0\\
	H^\mu_{N+2,1}(t) & 0 & \cdots & 0 & 0 & 0 & 0 & \cdots & 0 & 0\\
	\vdots & \vdots & \ddots & \vdots & \vdots & \vdots & \vdots & \ddots &\vdots & \vdots\\
	H^\mu_{2N-1,1}(t) & H^\mu_{2N-1,2}(t) & \cdots & 0 & 0 & 0 & 0 & \cdots & 0 & 0\\
	H^\mu_{2N,1}(t) & H^\mu_{2N,2}(t) & \cdots & H^\mu_{2N,N-1}(t) & 0 & 0 & 0 & \cdots & 0 & 0
\end{array}\right).~\label{HamMuMatr}
\end{aligned}
\end{eqnarray}
Here the block matrices $H^\mu_d(t)$ and $H_s^\mu(t)$ [$(H_s^\mu(t))^\dagger$] describe the eigenfrequencies of the bosonic modes and the squeezing interactions among them, respectively, and $\bf{0}$ is an $N\times N$ zero matrix. The off-diagonal matrix elements are related by complex conjugation as $H^\mu_{j,N+m}(t)=[H^\mu_{N+m,j}(t)]^*$ for $m\geq j+1$ with $j=1,2,\cdots,N-1$. For example, using the single-mode ancillary operators $[\hat{\mu}_1(t),\hat{\mu}_1^\dagger(t)]^T=\mathcal{M}(t)[\hat{a}_1(t),\hat{a}_1^\dagger(t)]^T$ with $\mathcal{M}(t)$ in Eq.~(\ref{Mone}), the single-mode squeezing Hamiltonian in Eq.~(\ref{HamNHsing}) in the absence of $\gamma$ can be expressed as
\begin{eqnarray}
&\hat{H}(t)=\begin{pmatrix}\hat{\mu}_1^\dagger(t)&\hat{\mu}_1(t)\end{pmatrix}
\left(\begin{array}{c|c}\omega\cosh^2\theta+\Omega\sinh2\theta\cos(\varphi+\alpha)
&H_{1,2}^\mu(t)\\
    \hline
   [H_{1,2}^\mu(t)]^*&\omega\sinh^2\theta+\Omega\sinh2\theta\cos(\varphi+\alpha)\end{array}\right)
    \begin{pmatrix}\hat{\mu}_1(t)\\ \hat{\mu}_1^\dagger(t)\end{pmatrix},
\end{eqnarray}
with $H_{1,2}^\mu(t)=[\omega\sinh\theta\cosh\theta+\Omega\cosh2\theta\cos(\varphi+\alpha)
+i\Omega\sin(\varphi+\alpha)]\exp{[-i\alpha(t)]}$. And using the two-mode ancillary operators $[\hat{\mu}_1(t),\hat{\mu}_2(t),\hat{\mu}_1^\dagger(t),\hat{\mu}_2^\dagger(t)]^T
=\mathcal{M}(t)[\hat{a}_1(t),\hat{a}_2(t),\hat{a}_1^\dagger(t),\hat{a}_2^\dagger(t)]$ with $\mathcal{M}(t)$ in Eq.~(\ref{AnciFull}), the two-mode Hamiltonian~(\ref{HamTwo}) with $\gamma=0$ can be written as
\begin{eqnarray}
\hat{H}(t)&=\begin{pmatrix}\hat{\mu}_1^\dagger(t) & \hat{\mu}_2^\dagger(t) &\hat{\mu}_1(t) &\hat{\mu}_2(t)\end{pmatrix}\left(\begin{array}{cc|cc}
H_{1,1}^\mu(t)& 0 & 0 & H_{1,4}^\mu(t)\\
    0& H_{2,2}^\mu(t) & H_{2,3}^\mu(t) & 0\\
    \hline
    0&{[H_{2,3}^\mu(t)]}^*&H_{2,2}^\mu(t)-\omega_2(t)&0\\
    {[H_{1,4}^\mu(t)]}^*& 0 & 0 & H_{1,1}^\mu(t)-\omega_1(t)
    \end{array}\right)\begin{pmatrix}\hat{\mu}_1(t)\\ \hat{\mu}_2(t)\\ \hat{\mu}_1^\dagger(t)\\ \hat{\mu}_2^\dagger(t)\end{pmatrix}
\end{eqnarray}
with $H_{1,1}^\mu(t)=\omega_1(t)\cosh^2\theta(t)+g(t)\sinh2\theta(t)\cos[\varphi+\alpha(t)]$, $H_{2,2}^\mu(t)=\omega_2(t)\cosh^2\theta(t)$, $H_{2,3}^\mu(t)=\omega_2(t)\sinh\theta(t)\cosh\theta(t)\exp[-i\alpha(t)]$, and $H_{1,4}^\mu(t)=[\omega_1(t)\sinh\theta(t)\cosh\theta(t)+g(t)\cosh2\theta(t)\cos(\varphi+\alpha(t))]\exp[-i\alpha(t)]$.

In the rotating frame with respect to $\mathcal{V}(t)$ in Eq.~(\ref{Unitrans}), the Hamiltonian~(\ref{HamMuAppend}) can be expressed in the stationary ancillary representation as
\begin{equation}\label{HamMuStaAppend}
\begin{aligned}
\hat{H}_{\rm rot}(t)&=\mathcal{V}^\dagger(t)\hat{H}(t)\mathcal{V}(t)-i\mathcal{V}^\dagger(t)
\frac{\partial\mathcal{V}(t)}{\partial t}=\begin{pmatrix}\vec{\mu}_0^\dagger&\vec{\mu}_0\end{pmatrix}\left[H^\mu(t)-\mathcal{A}(t)\right]
\begin{pmatrix}\vec{\mu}_0^T\\(\vec{\mu}_0^\dagger)^T\end{pmatrix}
=\begin{pmatrix}\vec{\mu}_0^\dagger&\vec{\mu}_0\end{pmatrix}\mathcal{H}(t)
\begin{pmatrix}\vec{\mu}_0^T\\(\vec{\mu}_0^\dagger)^T\end{pmatrix},
\end{aligned}
\end{equation}
where the gauge or holonomic component~\cite{Zhang2023Geometric} is given by $\mathcal{A}(t)=i[\mathcal{M}^\dagger(0)]^{-1}\mathcal{M}^{-1}(t)\dot{\mathcal{M}}(t)\mathcal{M}^{-1}(0)$ with the stationary symplectic matrix $\mathcal{M}(0)$. Note here the gauge potential $\mathcal{A}(t)$ is determined by the ancillary operators $\{\hat{\mu}_k(t)\}$ instead of by the eigenvalue equation~\cite{Michael2017Geometry}. The $2N\times 2N$ coefficient matrix $\mathcal{H}(t)$ reads
\begin{eqnarray}
\begin{aligned}
\mathcal{H}(t)&=\left(\begin{array}{c|c}\mathcal{H}_d(t)&\mathcal{H}_s(t)\\
    \hline
    \mathcal{H}_s^\dagger(t)&\bf{0}\end{array}\right)\\
    &=\left(\begin{array}{ccccc|ccccc}
	\mathcal{H}_{1,1}(t) & 0 & \cdots & 0 & 0 & 0 & \mathcal{H}_{1,N+2}(t) & \cdots & \mathcal{H}_{1,2N-1}(t) & \mathcal{H}_{1,2N}(t)\\
	0 & \mathcal{H}_{2,2}(t) & \cdots & 0 & 0 &0 & 0 & \cdots & \mathcal{H}_{2,2N-1}(t) & \mathcal{H}_{2,2N}(t)\\
	\vdots & \vdots & \ddots & \vdots & \vdots & \vdots & \vdots & \ddots &\vdots & \vdots\\
	0 & 0 & \cdots & \mathcal{H}_{N-1,N-1}(t) & 0 & 0 & 0 & \cdots & 0 & \mathcal{H}_{N-1,2N}(t)\\
	0 & 0 & \cdots & 0 & \mathcal{H}_{N,N}(t) & 0 & 0 & \cdots & 0 & 0\\
	\hline
	0 & 0 & \cdots & 0 & 0 & 0 & 0 & \cdots & 0 & 0\\
	\mathcal{H}_{N+2,1}(t) & 0 & \cdots & 0 & 0 & 0 & 0 & \cdots & 0 & 0\\
	\vdots & \vdots & \ddots & \vdots & \vdots & \vdots & \vdots & \ddots &\vdots & \vdots\\
	\mathcal{H}_{2N-1,1}(t) & \mathcal{H}_{2N-1,2}(t) & \cdots & 0 & 0 & 0 & 0 & \cdots & 0 & 0\\
	\mathcal{H}_{2N,1}(t) & \mathcal{H}_{2N,2}(t) & \cdots & \mathcal{H}_{2N,N-1}(t) & 0 & 0 & 0 & \cdots & 0 & 0
\end{array}\right),~\label{HamMuStaMatr}
\end{aligned}
\end{eqnarray}
where the diagonal terms can be used to define the gauge-invariant global phase $f_k(t)\equiv\int_0^t\mathcal{H}_{kk}(s)ds$, $1\leq k\leq N$.

In the following, we prove that the constraints about $\tilde{\mu}_k(t)$ and $\tilde{\mu}_k^\dagger(t)$ in Eq.~(\ref{Quanti}) by the Heisenberg equation~(\ref{vonNeu}) with the Hamiltonian~(\ref{HamMuAppend}) constitutes a necessary and sufficient condition for all the elements in the $k$th row and the $k$th column of both block $\mathcal{H}_s(t)$ and $\mathcal{H}_s^\dagger(t)$ in Eq.~(\ref{HamMuStaMatr}) vanish, where $k\in\{1,2,\cdots,N\}$.

\emph{Necessary condition.---} If the elements in the $N+k$th row and $k$th column and those in the $k$th row and $N+k$th column of $\mathcal{H}(t)$ in Eq.~(\ref{HamMuStaMatr}) vanish, then we have
\begin{equation}\label{HamMuStaDiaApp}
\begin{aligned}
\hat{H}_{\rm rot}(t)&=\sum_{j=1}^N\dot{f}_j(t)\hat{\mu}_j^\dagger(0)\hat{\mu}_j(0)+\sum_{j\neq k,j=1}^{N-1}\sum_{m\neq k, m\geq j+1}^N\Big[\mathcal{H}_{j,N+m}(t)\hat{\mu}_j^\dagger(0)\hat{\mu}_m^\dagger(0)\\
&+\mathcal{H}_{N+m,j}(t)\hat{\mu}_m(0)\hat{\mu}_j(0)\Big]
=\begin{pmatrix}\vec{\mu}_0^\dagger&\vec{\mu}_0\end{pmatrix}\mathcal{H}(t)
\begin{pmatrix}\vec{\mu}_0^T\\(\vec{\mu}_0^\dagger)^T\end{pmatrix}
\end{aligned}
\end{equation}
with the block coefficient matrix $\mathcal{H}(t)$ as
\begin{eqnarray}
\begin{aligned}
\mathcal{H}(t)&=\left(\begin{array}{ccccccc|ccccccc}
	\dot{f}_1 & 0 & \cdots & 0 &\cdots& 0 & 0 & 0 & \mathcal{H}_{1,N+2} & \cdots & 0 & \cdots & \mathcal{H}_{1,2N-1} & \mathcal{H}_{1,2N}\\
	0 & \dot{f}_2 & \cdots & 0 &\cdots& 0 & 0 & 0 & 0 & \cdots & 0 & \cdots & \mathcal{H}_{2,2N-1} & \mathcal{H}_{2,2N}\\
	\vdots & \vdots & \ddots & \vdots & \ddots & \vdots & \vdots & \vdots & \vdots & \ddots & \vdots & \ddots & \vdots & \vdots\\
	0 & 0 & \cdots & \dot{f}_{k} & \cdots & 0 & 0 & 0 & 0 & \cdots & 0 & \cdots & 0 & 0 \\
	\vdots & \vdots & \ddots& \vdots & \ddots & \vdots &\vdots & \vdots & \vdots & \ddots & \vdots & \ddots & \vdots & \vdots\\
	0 & 0 & \cdots & 0 & \cdots & \dot{f}_{N-1} & 0 & 0 & 0 & \cdots & 0 & \cdots & 0 & \mathcal{H}_{N-1,2N}\\
	0 & 0 & \cdots & 0 & \cdots & 0 & \dot{f}_N & 0 & 0 & \cdots & 0 & \cdots & 0 & 0\\
	\hline
	0 & 0 & \cdots & 0 & \cdots & 0 & 0 & 0 & 0 & \cdots & 0 & \cdots & 0 & 0\\
	\mathcal{H}_{N+2,1} & 0 & \cdots & 0 & \cdots & 0 & 0 & 0 & 0 & \cdots & 0 & \cdots & 0 & 0\\
	\vdots & \vdots & \ddots& \vdots & \ddots & \vdots &\vdots & \vdots & \vdots & \ddots & \vdots & \ddots & \vdots & \vdots\\
	0 & 0 & \cdots & 0 & \cdots & 0 & 0 & 0 & 0 & \cdots & 0 & \cdots & 0 & 0\\
	\vdots & \vdots & \ddots& \vdots & \ddots & \vdots &\vdots & \vdots & \vdots & \ddots & \vdots & \ddots & \vdots & \vdots\\
	\mathcal{H}_{2N-1,1} & \mathcal{H}_{2N-1,2} & \cdots & 0 & \cdots & 0 & 0 & 0 & 0 & \cdots & 0 & \cdots & 0 & 0\\
	\mathcal{H}_{2N,1} & \mathcal{H}_{2N,2} & \cdots & 0 & \cdots & \mathcal{H}_{2N,N-1} & 0 & 0 & 0 & \cdots & 0 & \cdots & 0 & 0
\end{array}\right),~\label{HamMuStaMatrBlock}
\end{aligned}
\end{eqnarray}
where the time-dependence of matrix elements is omitted for simplicity. It is straightforward to verify the commutation relation between $\hat{H}_{\rm rot}(t)$ in Eq.~(\ref{HamMuStaDiaApp}) and $\tilde{\mu}_k(0)$:
\begin{equation}\label{HamMuStaComm}
\begin{aligned}
-i\left[\hat{H}_{\rm rot}(t),\tilde{\mu}_k(0)\right]&=-i\Big[\sum_{j=1}^N\dot{f}_j(t)\hat{\mu}_j^\dagger(0)\hat{\mu}_j(0)+\sum_{j\neq k,j=1}^{N-1}\sum_{m\neq k, m\geq j+1}^N\Big[\mathcal{H}_{j,N+m}(t)\hat{\mu}_j^\dagger(0)\hat{\mu}_m^\dagger(0)\\
&+\mathcal{H}_{N+m,j}(t)\hat{\mu}_m(0)\hat{\mu}_j(0)\Big], e^{if_k(t)}\hat{\mu}_k(0)\Big]
=i\dot{f}_k(t)e^{if_k(t)}\hat{\mu}_k(0)=\frac{\partial[e^{if_k(t)}\hat{\mu}_k(0)]}{\partial t}=\frac{\partial\tilde{\mu}_k(0)}{\partial t}.
\end{aligned}
\end{equation}
The preceding derivation also applies to $\tilde{\mu}_k^\dagger(0)$.

Using Eqs.~(\ref{Unitrans}) and (\ref{HamMuSta}), Eq.~(\ref{HamMuStaComm}) is equivalent to
\begin{equation}\label{HamCommTra}
-i\left[\mathcal{V}^\dagger(t)\hat{H}(t)\mathcal{V}(t)-i\mathcal{V}^\dagger(t)
\frac{\partial\mathcal{V}(t)}{\partial t}, \mathcal{V}^\dagger(t)\mathcal{J}_k(t)\mathcal{V}(t)\right]
=\frac{\partial\left[\mathcal{V}^\dagger(t)\mathcal{J}_k(t)\mathcal{V}(t)\right]}{\partial t},
\end{equation}
with $\mathcal{J}_k(t)=\tilde{\mu}_k(t)$ and $\tilde{\mu}_k^\dagger(t)$. And it expands as
\begin{equation}\label{HamCommTraSim}
\begin{aligned}
\mathcal{V}^\dagger(t)\frac{\partial\mathcal{J}_k(t)}{\partial t}\mathcal{V}(t)&=-i\mathcal{V}^\dagger(t)\left[\hat{H}(t), \mathcal{J}_k(t)\right]\mathcal{V}(t)
-\left[\mathcal{V}^\dagger(t)\frac{\partial\mathcal{V}(t)}{\partial t}\mathcal{V}^\dagger(t)\mathcal{J}_k(t)\mathcal{V}(t)
-\mathcal{V}^\dagger(t)\mathcal{J}_k(t)\frac{\partial\mathcal{V}(t)}{\partial t}\right]\\
&-\left[\frac{\partial\mathcal{V}^\dagger(t)}{\partial t}\mathcal{J}_k(t)\mathcal{V}(t)+\mathcal{V}^\dagger(t)\mathcal{J}_k(t)\frac{\partial\mathcal{V}(t)}{\partial t}\right].
\end{aligned}
\end{equation}
By left-multiplying Eq.~(\ref{HamCommTraSim}) with $\mathcal{V}(t)$ and right-multiplying with $\mathcal{V}^\dagger(t)$ and due to the fact that $\partial_t[\mathcal{V}^\dagger(t)\mathcal{V}(t)]=\partial_t\mathcal{V}^\dagger(t)\mathcal{V}(t)
+\mathcal{V}^\dagger(t)\partial_t\mathcal{V}(t)=0$, we have
\begin{equation}\label{vonNeuAppend}
\frac{\partial\mathcal{J}_k(t)}{\partial t}=-i\left[\hat{H}(t),\mathcal{J}_k(t)\right],
\end{equation}
which is exactly the same as Eq.~(\ref{vonNeu}).

\emph{Sufficient condition.---} We start from the Heisenberg equation~(\ref{vonNeu}) or Eq.~(\ref{vonNeuAppend}) in the time-dependent representation, and rotate it back to the stationary representation via the inverse transformations of Eqs.~(\ref{Unitrans}) and (\ref{HamMuSta}). Particularly, Eq.~(\ref{vonNeu}) or Eq.~(\ref{vonNeuAppend}) is transformed as
\begin{equation}\label{vonNeuTurn}
\frac{\partial[\mathcal{V}(t)\mathcal{J}_k(0)\mathcal{V}^\dagger(t)]}{\partial t}=-i\left[\mathcal{V}(t)\hat{H}_{\rm rot}(t)\mathcal{V}^\dagger(t)+i\frac{\partial \mathcal{V}(t)}{\partial t}\mathcal{V}^\dagger(t),\mathcal{V}(t)\mathcal{J}_k(0)\mathcal{V}^\dagger(t)\right],
\end{equation}
where $\mathcal{J}_k(0)=\tilde{\mu}_k(0)\equiv\exp[if_k(t)]\hat{\mu}_k(0)$ and $\tilde{\mu}_k^\dagger(0)\equiv\exp[-if_k(t)]\hat{\mu}_k^\dagger(0)$. Using the chain rule, we have
\begin{equation}\label{vonNeuChain}
\begin{aligned}
\mathcal{V}(t)\frac{\partial\mathcal{J}_k(0)}{\partial t}\mathcal{V}^\dagger(t)&=-i\mathcal{V}(t)\left[\hat{H}_{\rm rot}(t),\mathcal{J}_k(0)\right]\mathcal{V}^\dagger(t)+\left[\frac{\partial\mathcal{V}(t)}{\partial t}\mathcal{J}_k(0)\mathcal{V}^\dagger(t)-\mathcal{V}(t)\mathcal{J}_k(0)\mathcal{V}^\dagger(t)
\frac{\partial\mathcal{V}(t)}{\partial t}\mathcal{V}^\dagger(t)\right]\\
&-\frac{\partial\mathcal{V}(t)}{\partial t}\mathcal{J}_k(0)\mathcal{V}^\dagger(t)-\mathcal{V}(t)\mathcal{J}_k(0)\frac{\partial\mathcal{V}^\dagger(t)}{\partial t}.
\end{aligned}
\end{equation}
Upon left-multiplying Eq.~(\ref{vonNeuChain}) with $\mathcal{V}^\dagger(t)$ and right-multiplying it with $\mathcal{V}(t)$, it returns to Eq.~(\ref{HamMuStaComm}):
\begin{equation}\label{vonNeuSta}
\frac{\partial\mathcal{J}_k(0)}{\partial t}=-i\left[\hat{H}_{\rm rot}(t),\mathcal{J}_k(0)\right].
\end{equation}
Equation~(\ref{HamMuStaComm}) is valid if and only if $\hat{H}_{\rm rot}(t)$ takes the form in Eq.~(\ref{HamMuStaDiaApp}), since any extra interaction Hamiltonian involving the $k$th stationary ancillary operators, i.e., $\hat{\mu}_k^\dagger(0)$ and $\hat{\mu}_k(0)$, would violate the commutation relation. For example, when Eq.~(\ref{HamMuStaDiaApp}) becomes
\begin{equation}\label{HamMuStaDiaExt}
\hat{H}_{\rm rot}'(t)=\hat{H}_{\rm rot}(t)+\mathcal{H}_{k',N+k}(t)\hat{\mu}_{k'}^\dagger(0)\hat{\mu}_{k}^\dagger(0)
+\mathcal{H}_{N+k,k'}(t)\hat{\mu}_{k}(0)\hat{\mu}_{k'}(0),
\end{equation}
where $\mathcal{H}_{k',N+k}(t)=[\mathcal{H}_{N+k,k'}(t)]^*$. Then the commutation relation between $\hat{H}_{\rm rot}'(t)$ and $\hat{\mu}_k(0)$ is found to be
\begin{equation}\label{HamMuStaCommExt}
-i\left[\hat{H}_{\rm rot}'(t), \tilde{\mu}_k(0)\right]
=ie^{if_k(t)}\left[\dot{f}_k(t)\hat{\mu}_k(0)+\mathcal{H}_{k',N+k}(t)\hat{\mu}_{k'}^\dagger(0)\right]
\neq\frac{\partial\tilde{\mu}_k(0)}{\partial t}.
\end{equation}
This failure applies to arbitrary $k'$ and $\tilde{\mu}_k^\dagger(0)$. Then we complete the proof about the sufficient condition.

With $\hat{H}_{\rm rot}(t)$ in Eq.~(\ref{HamMuStaDiaApp}), the stationary ancillary operators $\hat{\mu}_k(0)$ and $\hat{\mu}_k^\dagger(0)$ are decoupled from all other operators, i.e., $\hat{\mu}_j(0)$ and $\hat{\mu}_j^\dagger(0)$ for $j\neq k$. Then by the Dyson series~\cite{Dyson1949TheRadiation} or the Magnus expansion~\cite{Blanes2009Magnus}, the time evolution operator in the stationary ancillary representation can be written as
\begin{equation}\label{UStaDia}
\hat{U}_{\rm rot}(t)=e^{-i\left[\sum_{j=1}^N\sum_{m\geq j+1}^Nh_{jm}(t)\hat{\mu}_j^\dagger(0)\hat{\mu}_m^\dagger(0)+{\rm H.c.}\right]}e^{-if_k(t)\hat{\mu}_k^\dagger(0)\hat{\mu}_k(0)},\quad j,m\neq k.
\end{equation}
Despite the elements $h_{jm}(t)$ are generally hard to be evaluated, they are not relevant to the exact dynamics of the system once the initial state is fully determined by $\hat{\mu}_k(0)$. By the Heisenberg equation with $\hat{U}_{\rm rot}(t)$ in Eq.~(\ref{UStaDia}), the dynamics of the ancillary operators $\hat{\mu}_k(0)$ and $\hat{\mu}_k^\dagger(0)$ are found to be
\begin{equation}\label{HermDyn}
\begin{aligned}
\hat{v}_k(t)&=\hat{U}_{\rm rot}^\dagger(t)\hat{\mu}_k(0)\hat{U}_{\rm rot}(t)=e^{-if_k(t)}\hat{\mu}_k(0),\\
\hat{v}_k^\dagger(t)&=\hat{U}_{\rm rot}^\dagger(t)\hat{\mu}_k^\dagger(0)\hat{U}_{\rm rot}(t)=e^{if_k(t)}\hat{\mu}_k^\dagger(0),
\end{aligned}
\end{equation}
According to Eqs.~(\ref{Unitrans}) and (\ref{HamMuSta}), the dynamics of $\hat{\mu}_k(0)$ and $\hat{\mu}_k^\dagger(0)$ in the original picture can be obtained as
\begin{equation}\label{HermDynOriDer}
\begin{aligned}
\hat{\mu}_k(0)&\rightarrow\mathcal{V}(t)\hat{v}_k(t)\mathcal{V}^\dagger(t)=e^{-if_k(t)}\hat{\mu}_k(t),\\
\hat{\mu}_k^\dagger(0)&\rightarrow\mathcal{V}(t)\hat{v}_k^\dagger(t)\mathcal{V}^\dagger(t)=e^{if_k(t)}\hat{\mu}_k^\dagger(t),
\end{aligned}
\end{equation}
which takes the same form as Eq.~(\ref{HermDynOri}).

\subsection{Nonlinear Hamiltonian}\label{nonlinear}

This section demonstrates that the Heisenberg equation~(\ref{vonNeu}) for our universal quantum control also applies to the systems governed by more nonlinear Hamiltonian, such as the multi-photon spontaneous parametric down-conversion~\cite{Maria2011Nonlinear,Borshchevskaya2015Three,Akbari2016Third,Chang2020Observation}, thereby yielding the dynamical invariants, which do not necessarily satisfy the canonical commutation relation, for nonadiabatic passages.

Consider a general two-mode bosonic system governed by the nonlinear Hamiltonian
\begin{equation}\label{HamNonLine}
\hat{H}(t)=\omega_1\hat{a}_1^{\dagger}\hat{a}_1+\omega_2\hat{a}_2^{\dagger}\hat{a}_2
+g(t)\left[e^{i\varphi}\left(\hat{a}_1^\dagger\right)^j\hat{a}_2^k
+e^{-i\varphi}\hat{a}_1^j\left(\hat{a}_2^\dagger\right)^k\right],
\end{equation}
where $\omega_1$ and $\omega_2$ are the eigen-frequencies of the two modes, $g(t)$ and $\varphi$ are the coupling strength and the relative phase between them, respectively, and $j$ and $k$ are arbitrary nonnegative integers. In the rotating frame with respect to the free Hamiltonian $\hat{H}_0=\omega_1\hat{a}_1^{\dagger}\hat{a}_1+\omega_2\hat{a}_2^{\dagger}\hat{a}_2$, the Hamiltonian in the interaction picture can be obtained as
\begin{equation}\label{HamNonLineInt}
\hat{H}_I(t)=g(t)\left[e^{-i(j\omega_1-k\omega_2)t+i\varphi}\left(\hat{a}_1^\dagger\right)^j\hat{a}_2^k
+e^{i(j\omega_1-k\omega_2)t-i\varphi}\hat{a}_1^j\left(\hat{a}_2^\dagger\right)^k\right].
\end{equation}
Under the multi-photon resonant condition $j\omega_1=k\omega_2$ and the condition of $\varphi=\pi/2$, Eq.~(\ref{HamNonLineInt}) can be reduced as
\begin{equation}\label{HamNonLineIntReson}
\hat{H}_I(t)=ig(t)\left[\left(\hat{a}_1^\dagger\right)^j\hat{a}_2^k-\hat{a}_1^j\left(\hat{a}_2^\dagger\right)^k\right].
\end{equation}
In the Fock state basis, it is written as
\begin{equation}\label{HamNonLineSum}
\hat{H}_I(t)=ig(t)\sum_{N=k}^\infty\sum_{n=0}^{N-k}\mathcal{N}_n\mathcal{N}_{N,n}\left[|n+j,N-n-k\rangle\langle n,N-n|-|n,N-n\rangle\langle n+j,N-n-k|\right],
\end{equation}
where the coefficients $\mathcal{N}_n\equiv\sqrt{(n+j)!/n!}$ and $\mathcal{N}_{N,n}\equiv\sqrt{(N-n)!/(N-n-k)!}$, and $|n,m\rangle=|n\rangle_1\otimes|m\rangle_2$ with $|n\rangle_1$ and $|m\rangle_2$ denoting the Fock states of modes $a_1$ and $a_2$, respectively.

The dynamical invariant of this system can be assumed in the form
\begin{equation}\label{invariant}
\hat{\mu}(t)=\sum_{n=j}^\infty\sum_{m=0}^\infty\cos\theta_{n,m}(t)|n-j,m\rangle\langle n,m|-\sum_{n=0}^\infty\sum_{m=k}^\infty\sin\tilde{\theta}_{n,m}(t)|n,m-k\rangle\langle n,m|
\end{equation}
with the time-dependent parameters $\theta_{n,m}(t)$ and $\tilde{\theta}_{n,m}(t)$. Substitute the Hamiltonian $\hat{H}_I(t)$ in Eq.~(\ref{HamNonLineIntReson}) or (\ref{HamNonLineSum}) and $\hat{\mu}(t)$ in Eq.~(\ref{invariant}) into the Heisenberg equation (\ref{vonNeu}) or (\ref{vonNeuAppend}), one can obtain the constraints
\begin{equation}\label{constrNonline}
\begin{aligned}
&\dot{\tilde{\theta}}_{n+j,N-n}\cos\tilde{\theta}_{n+j,N-n}=-g(t)\mathcal{N}_{N,n}
\left(\mathcal{N}_n\cos\theta_{n+j,N-n}-\mathcal{N}_{n+j}\cos\theta_{n+2j,N-n-k}\right),\quad N\ge k, \quad 0\leq n\leq N-k,\\
&\dot{\tilde{\theta}}_{n,N-n}\cos\tilde{\theta}_{n,N-n}=g(t)\mathcal{N}_n\mathcal{N}_{N,n}\cos\theta_{n+j,N-n-k}, \quad N\ge n+k, \quad 0\leq n\leq j-1,\\
&\dot{\theta}_{n+j,N-n}\sin\theta_{n+j,N-n}
=-g(t)\mathcal{N}_n\left(\mathcal{N}_{N,n}\sin\tilde{\theta}_{n+j,N-n}-\mathcal{N}_{N+k,n}\sin\tilde{\theta}_{n,N-n+k}\right),
\quad N\ge k, \quad 0\leq n\leq N-k,\\
&\dot{\theta}_{n+j,N-n-k}\sin\theta_{n+j,N-n-k}=g(t)\mathcal{N}_n\mathcal{N}_{N,n}\sin\tilde{\theta}_{n,N-n},\quad N\ge n+k, \quad n\ge0,
\end{aligned}
\end{equation}
with $\mathcal{N}_n\cos\theta_{n+2j,N-n-k}=\mathcal{N}_{n+j}\cos\theta_{n+j,N-n}$ and $\mathcal{N}_{N,n}\sin\tilde{\theta}_{n,N-n+k}=\mathcal{N}_{N+k,n}\sin\tilde{\theta}_{n+j,N-n}$. In other words, for the system dynamics governed by $\hat{H}_I(t)$ in Eq.~(\ref{HamNonLineIntReson}) or (\ref{HamNonLineSum}), the dynamical invariant defined in Eq.~(\ref{invariant}) can be directly exploited to control the system trajectories subject to the constraints in Eq.~(\ref{constrNonline}).

As an example, one can consider the three-photon parametric down-conversion governed by the Hamiltonian~(\ref{HamNonLineSum}) with $j=1$ and $k=2$, i.e.,
\begin{equation}\label{HamNonLineSumRedu}
\hat{H}_I(t)=ig(t)\sum_{N=2}^\infty\sum_{n=0}^{N-2}\mathcal{N}_n\mathcal{N}_{N,n}\left[|n+1,N-n-2\rangle\langle n,N-n|-|n,N-n\rangle\langle n+1,N-n-2|\right]
\end{equation}
with $\mathcal{N}_n=\sqrt{(n+1)!/n!}$ and $\mathcal{N}_{N,n}=\sqrt{(N-n)!/(N-n-2)!}$. The relevant dynamical invariant in Eq.~(\ref{invariant}) is
\begin{equation}\label{invariantRed}
\hat{\mu}(t)=\sum_{n=1}^\infty\sum_{m=0}^\infty\cos\theta_{n,m}(t)|n-1,m\rangle\langle n,m|-\sum_{n=0}^\infty\sum_{m=2}^\infty\sin\tilde{\theta}_{n,m}(t)|n,m-2\rangle\langle n,m|.
\end{equation}
Then the constraint conditions in Eq.~(\ref{constrNonline}) become
\begin{equation}\label{constrNonlineRed}
\begin{aligned}
&\dot{\tilde{\theta}}_{n+1,N-n}\cos\tilde{\theta}_{n+1,N-n}
=-g(t)\mathcal{N}_{N,n}\left(\mathcal{N}_n\cos\theta_{n+1,N-n}-\mathcal{N}_{n+1}\cos\theta_{n+2,N-n-2}\right), \quad N\ge 2, \quad 0\leq n\leq N-2,\\
&\dot{\tilde{\theta}}_{n,N-n}\cos\tilde{\theta}_{n,N-n}=g(t)\mathcal{N}_n\mathcal{N}_{N,n}\cos\theta_{n+1,N-n-2},\quad N\ge n+2,\quad n=0,\\
&\dot{\theta}_{n+1,N-n}\sin\theta_{n+1,N-n}
=-g(t)\mathcal{N}_n\left(\mathcal{N}_{N,n}\sin\tilde{\theta}_{n+1,N-n}-\mathcal{N}_{N+2,n}\sin\tilde{\theta}_{n,N-n+2}\right),
\quad N\ge 2,\quad 0\leq n\leq N-2,\\
&\dot{\theta}_{n+1,N-n-2}\sin\theta_{n+1,N-n-2}=g(t)\mathcal{N}_n\mathcal{N}_{N,n}\sin\tilde{\theta}_{n,N-n},\quad N\ge n+2, \quad n\ge0
\end{aligned}
\end{equation}
with $\mathcal{N}_n\cos\theta_{n+2,N-n-2}=\mathcal{N}_{n+1}\cos\theta_{n+1,N-n}$ and $\mathcal{N}_{N,n}\sin\tilde{\theta}_{n,N-n+2}=\mathcal{N}_{N+2,n}\sin\tilde{\theta}_{n+1,N-n}$. Without loss of generality, one can consider a target that the system prepared as $|\psi(0)\rangle=|3,0\rangle$ evolves to the final state $|\psi(\tau)\rangle=|0,6\rangle$ at time $t=\tau$. In accordance to Eq.~(\ref{invariantRed}) and its Hermitian conjugate, under the boundary conditions $\cos\theta_{4,0}(0)=k\pi$ and $\sin\tilde{\theta}_{0,8}(\tau)=k\pi+\pi/2$ with $k\in Z$, the effective time-evolution operator at the desired moment is given by $U_{\rm eff}(\tau)=|0,6\rangle\langle3,0|+|3,0\rangle\langle6,0|$. Consequently, the target state can be obtained as $|\psi(\tau)\rangle=U_{\rm eff}(\tau)|\psi(0)\rangle=|0,6\rangle$.

\section{Universal quantum control with time-dependent non-Hermitian Hamiltonian}\label{SufNecDiag}

This section demonstrates that when only $\tilde{\mu}_k^\dagger(t)$ but not $\tilde{\mu}_k(t)$ in Eq.~(\ref{Quanti}) satisfy the Heisenberg equation~(\ref{vonNeu}) with the time-dependent non-Hermitian Hamiltonian $\hat{H}(t)\neq\hat{H}^\dagger(t)$, it suffices to demonstrate the system dynamics in the ket space described by Eq.~(\ref{NonHermDynOri}).

Under the biorthogonal assumption~\cite{Brody2013Biorhogonal}, the dynamics of non-Hermitian systems can be described by two sets of time-dependent Schr\"odinger equations as
\begin{equation}\label{SchAPP}
i\frac{\partial}{\partial t}|\psi(t)\rangle=\hat{H}(t)|\psi(t)\rangle,\quad i\frac{\partial}{\partial t}|\phi(t)\rangle=\hat{H}^\dagger(t)|\phi(t)\rangle
\end{equation}
with the pure-state solutions $|\psi(t)\rangle$ and $\langle\phi(t)|$ in the ket and bra spaces, respectively. The time-dependent non-Hermitian quadratic Hamiltonian $\hat{H}(t)$ can be written in the general form,
\begin{equation}\label{HamNonAPP}
\hat{H}(t)=\begin{pmatrix}\vec{a}^\dagger&\vec{a}\end{pmatrix}H^a(t)
\begin{pmatrix}\vec{a}^T\\(\vec{a}^\dagger)^T\end{pmatrix},
\end{equation}
where the $2N\times2N$ coefficient matrix $H^a(t)$ is not Hermitian, i.e., $H^a(t)\neq[H^a(t)]^\dagger$. The non-Hermitian Hamiltonian can reduce to the pseudo-Hermitian Hamiltonian when the spectrum is real, or when $|\psi(t)\rangle$ and $|\phi(t)\rangle$ share the same complex eigenvalues~\cite{Mosta2002Pseudo}.

By Eq.~(\ref{TransM}), the non-Hermitian Hamiltonian~(\ref{HamNonAPP}) can be expressed in the time-dependent ancillary representation as
\begin{equation}\label{HamNonHermMu}
\begin{aligned}
\hat{H}(t)&=\begin{pmatrix}\vec{\mu}^\dagger&\vec{\mu}\end{pmatrix}(\mathcal{M}^\dagger(t))^{-1}H^a(t)\mathcal{M}^{-1}(t)
\begin{pmatrix}\vec{\mu}^T\\(\vec{\mu}^\dagger)^T\end{pmatrix}
=\begin{pmatrix}\vec{\mu}^\dagger&\vec{\mu}\end{pmatrix}H^\mu(t)\begin{pmatrix}\vec{\mu}^T\\(\vec{\mu}^\dagger)^T\end{pmatrix},
\end{aligned}
\end{equation}
with the $2N\times2N$ rotated coefficient matrix $H^\mu(t)$. Also, in the rotating frame with respect to $\mathcal{V}(t)$ defined in Eq.~(\ref{Unitrans}), the transformed Hamiltonian can be expressed
\begin{equation}\label{HamMuStaNon}
\begin{aligned}
&\hat{H}_{\rm rot}(t)=\mathcal{V}^\dagger(t)\hat{H}(t)\mathcal{V}(t)-i\mathcal{V}^\dagger(t)\frac{\partial\mathcal{V}(t)}{\partial t}\\
&=\begin{pmatrix}\vec{\mu}_0^\dagger&\vec{\mu}_0\end{pmatrix}
\left[H^\mu(t)-\mathcal{A}(t)\right]\begin{pmatrix}\vec{\mu}_0^T\\(\vec{\mu}_0^\dagger)^T\end{pmatrix}
\equiv\begin{pmatrix}\vec{\mu}_0^\dagger&\vec{\mu}_0\end{pmatrix}\mathcal{H}(t)
\begin{pmatrix}\vec{\mu}_0^T\\(\vec{\mu}_0^\dagger)^T\end{pmatrix},
\end{aligned}
\end{equation}
where $\mathcal{H}(t)$ is a $2N\times2N$ coefficient matrix in the representation of stationary ancillary operators. Without loss of generality, we consider the bosonic modes coupled by the squeezing interaction. Thus the Hamiltonian in Eqs.~(\ref{HamNonHermMu}) and (\ref{HamMuStaNon}) take the same form as those in Eqs.~(\ref{HamMuAppend}) and~(\ref{HamMuStaAppend}) for the Hermitian case. However, here the coefficient matrices are non-Hermitian, i.e., $H^{\mu}(t)\neq[H^{\mu}(t)]^\dagger$ and $\mathcal{H}(t)\neq\mathcal{H}^\dagger(t)$. Consequently, the ancillary operator and its Hermitian conjugate cannot satisfy the Heisenberg equation~(\ref{vonNeu}) at the same time.

In the following, we prove that the Heisenberg equation~(\ref{vonNeu}) for $\tilde{\mu}_k^\dagger(t)$ with $\hat{H}(t)$~(\ref{HamNonAPP}) is a necessary and sufficient condition for all the elements in the $N+k$th row and $k$th column of the coefficient matrix $\mathcal{H}(t)$ in Eq.~(\ref{HamMuStaMatr}) to vanish, where $k\in\{1,2,\cdots,N\}$.

\emph{Necessary condition.---} If all the elements of the Hamiltonian $\hat{H}_{\rm rot}(t)$ in Eq.~(\ref{HamMuStaNon}) in the $N+k$th row and $k$th column are zero, then we have
\begin{equation}\label{HamMuStaNonHerm}
\begin{aligned}
&\hat{H}_{\rm rot}(t)=\sum_{j=1}^N\dot{f}_j(t)\hat{\mu}_j^\dagger(0)\hat{\mu}_j(0)+\sum_{j=1}^{N-1}\sum_{m\geq j+1}^{N}\mathcal{H}_{j,N+m}(t)\hat{\mu}_j^\dagger(0)\hat{\mu}_m^\dagger(0)\\
&+\sum_{j\neq k,j=1}^{N-1}\sum_{m\neq k, m\geq j+1}^N\mathcal{H}_{N+m,j}(t)\hat{\mu}_m(0)\hat{\mu}_j(0)
=\begin{pmatrix}\vec{\mu}_0^\dagger&\vec{\mu}_0\end{pmatrix}\mathcal{H}(t)
\begin{pmatrix}\vec{\mu}_0^T\\(\vec{\mu}_0^\dagger)^T\end{pmatrix},
\end{aligned}
\end{equation}
where the coefficient matrix $\mathcal{H}(t)$ can still be divided into four $N\times N$ blocks as
\begin{eqnarray}
\begin{aligned}
\mathcal{H}(t)&=\left(\begin{array}{ccccccc|ccccccc}
	\dot{f}_1 & 0 & \cdots & 0 &\cdots& 0 & 0 & 0 & \mathcal{H}_{1,N+2} & \cdots & \mathcal{H}_{1,N+3} & \cdots & \mathcal{H}_{1,2N-1} & \mathcal{H}_{1,2N}\\
	0 & \dot{f}_2 & \cdots & 0 &\cdots& 0 & 0 & 0 & 0 & \cdots & \mathcal{H}_{2,N+3} & \cdots & \mathcal{H}_{2,2N-1} & \mathcal{H}_{2,2N}\\
	\vdots & \vdots & \ddots & \vdots & \ddots & \vdots & \vdots & \vdots & \vdots & \ddots & \vdots & \ddots & \vdots & \vdots\\
	0 & 0 & \cdots & \dot{f}_{k} & \cdots & 0 & 0 & 0 & 0 & \cdots & 0 & \cdots & \mathcal{H}_{k,2N-1} & \mathcal{H}_{k,2N}\\
	\vdots & \vdots & \ddots& \vdots & \ddots & \vdots &\vdots & \vdots & \vdots & \ddots & \vdots & \ddots & \vdots & \vdots\\
	0 & 0 & \cdots & 0 & \cdots & \dot{f}_{N-1} & 0 & 0 & 0 & \cdots & 0 & \cdots & 0 & \mathcal{H}_{N-1,2N}\\
	0 & 0 & \cdots & 0 & \cdots & 0 & \dot{f}_N & 0 & 0 & \cdots & 0 & \cdots & 0 & 0\\
	\hline
    0 & 0 & \cdots & 0 & \cdots & 0 & 0 & 0 & 0 & \cdots & 0 & \cdots & 0 & 0\\
	\mathcal{H}_{N+2,1} & 0 & \cdots & 0 & \cdots & 0 & 0 & 0 & 0 & \cdots & 0 & \cdots & 0 & 0\\
	\vdots & \vdots & \ddots& \vdots & \ddots & \vdots &\vdots & \vdots & \vdots & \ddots & \vdots & \ddots & \vdots & \vdots\\
	0 & 0 & \cdots & 0 & \cdots & 0 & 0 & 0 & 0 & \cdots & 0 & \cdots & 0 & 0\\
	\vdots & \vdots & \ddots& \vdots & \ddots & \vdots &\vdots & \vdots & \vdots & \ddots & \vdots & \ddots & \vdots & \vdots\\
	\mathcal{H}_{2N-1,1} & \mathcal{H}_{2N-1,2} & \cdots & 0 & \cdots & 0 & 0 & 0 & 0 & \cdots & 0 & \cdots & 0 & 0\\
	\mathcal{H}_{2N,1} & \mathcal{H}_{2N,2} & \cdots & 0 & \cdots & \mathcal{H}_{2N,N-1} & 0 & 0 & 0 & \cdots & 0 & \cdots & 0 & 0
\end{array}\right),~\label{HamMuBlockNon}
\end{aligned}
\end{eqnarray}
The commutation relation between $\hat{H}_{\rm rot}(t)$ and $\tilde{\mu}_k^\dagger(0)$ is found to be
\begin{equation}\label{HamMuStaNonHermComm}
\begin{aligned}
&-i\Big[\hat{H}_{\rm rot}(t),\tilde{\mu}_k^\dagger(0)\Big]=-i\Big[\sum_{k=1}^N\dot{f}_k(t)\hat{\mu}_k^\dagger(0)\hat{\mu}_k(0)+\sum_{j=1}^{N-1}\sum_{m\geq j+1}^{N}\mathcal{H}_{j,N+m}(t)\hat{\mu}_j^\dagger(0)\hat{\mu}_m^\dagger(0)\\
&+\sum_{j\neq k,j=1}^{N-1}\sum_{m\neq k,m\geq j+1}^N\mathcal{H}_{N+m,j}(t)\hat{\mu}_m(0)\hat{\mu}_j(0),e^{-if_k(t)}\hat{\mu}_k^\dagger(0)\Big]=-i\dot{f}_k(t)e^{-if_k(t)}\hat{\mu}_k^\dagger(0)=\frac{\partial[e^{-if_k(t)}\hat{\mu}_k^\dagger(0)]}{\partial t}=\frac{\partial\tilde{\mu}_k^\dagger(0)}{\partial t}.
\end{aligned}
\end{equation}

Under Eqs.~(\ref{Unitrans}) and (\ref{HamMuSta}), Eq.~(\ref{HamMuStaNonHermComm}) can be transformed as
\begin{equation}\label{HamCommTraNon}
\frac{\partial\Big[\mathcal{V}^\dagger(t)\tilde{\mu}_k^\dagger(t)\mathcal{V}(t)\Big]}{\partial t}=-i\Big[\mathcal{V}^\dagger(t)\hat{H}(t)\mathcal{V}(t)-i\mathcal{V}^\dagger(t)
\frac{\partial\mathcal{V}(t)}{\partial t},\mathcal{V}^\dagger(t)\tilde{\mu}_k^\dagger(t)\mathcal{V}(t)\Big],
\end{equation}
which leads to
\begin{equation}\label{HamCommTraSimNon}
\begin{aligned}
\mathcal{V}^\dagger(t)\frac{\partial\tilde{\mu}_k^\dagger(t)}{\partial t}\mathcal{V}(t)&=-i\mathcal{V}^\dagger(t)
\left[\hat{H}(t),\tilde{\mu}_k^\dagger(t)\right]\mathcal{V}(t)
-\left[\mathcal{V}^\dagger(t)\frac{\partial\mathcal{V}(t)}{\partial t}\mathcal{V}^\dagger(t)\tilde{\mu}_k^\dagger(t)\mathcal{V}(t)
-\mathcal{V}^\dagger(t)\tilde{\mu}_k^\dagger(t)\frac{\partial\mathcal{V}(t)}{\partial t}\right]\\
&-\left[\frac{\partial\mathcal{V}^\dagger(t)}{\partial t}\tilde{\mu}_k^\dagger(t)\mathcal{V}(t)
+\mathcal{V}^\dagger(t)\tilde{\mu}_k^\dagger(t)\frac{\partial\mathcal{V}(t)}{\partial t}\right].
\end{aligned}
\end{equation}
Multiplying Eq.~(\ref{HamCommTraSimNon}) by $\mathcal{V}(t)$ from the left and by $\mathcal{V}^\dagger(t)$ from the right, we have
\begin{equation}\label{vonNeuAppendNon}
\frac{\partial\tilde{\mu}_k^\dagger(t)}{\partial t}=-i\left[\hat{H}(t),\tilde{\mu}_k^\dagger(t)\right],
\end{equation}
which takes the same form as Eq.~(\ref{vonNeu}) with $\mathcal{J}_k(t)=\tilde{\mu}_k^\dagger(t)$.

\emph{Sufficient condition.---} Consider the Heisenberg equation (\ref{vonNeu}) with $\mathcal{J}_k(t)=\tilde{\mu}_k^\dagger(t)$ or Eq.~(\ref{vonNeuAppendNon}), and rotate it to the stationary representation via Eqs.~(\ref{Unitrans}) and (\ref{HamMuSta}). We have
\begin{equation}\label{vonNeuTurnNon}
\frac{\partial[\mathcal{V}(t)\tilde{\mu}_k^\dagger(0)\mathcal{V}^\dagger(t)]}{\partial t}
=-i\left[\mathcal{V}(t)\hat{H}_{\rm rot}(t)\mathcal{V}^\dagger(t)+i\frac{\partial \mathcal{V}(t)}{\partial t}\mathcal{V}^\dagger(t),\mathcal{V}(t)\tilde{\mu}_k^\dagger(0)\mathcal{V}^\dagger(t)\right].
\end{equation}
By the chain rule, it turns out to be
\begin{equation}\label{vonNeuChainNon}
\begin{aligned}
\mathcal{V}(t)\frac{\partial\tilde{\mu}_k^\dagger(0)}{\partial t}\mathcal{V}^\dagger(t)&=-i\mathcal{V}(t)\left[\hat{H}_{\rm rot}(t), \tilde{\mu}_k^\dagger(0)\right]\mathcal{V}^\dagger(t)
+\left[\frac{\partial\mathcal{V}(t)}{\partial t}\tilde{\mu}_k^\dagger(0)\mathcal{V}^\dagger(t)
-\mathcal{V}(t)\tilde{\mu}_k^\dagger(0)\mathcal{V}^\dagger(t)\frac{\partial\mathcal{V}(t)}{\partial t}\mathcal{V}^\dagger(t)\right]\\
&-\frac{\partial\mathcal{V}(t)}{\partial t}\tilde{\mu}_k^\dagger(0)\mathcal{V}^\dagger(t)
-\mathcal{V}(t)\tilde{\mu}_k^\dagger(0)\frac{\partial\mathcal{V}^\dagger(t)}{\partial t}.
\end{aligned}
\end{equation}
Then by left-multiplying Eq.~(\ref{vonNeuChainNon}) with $\mathcal{V}^\dagger(t)$ and right-multiplying it with $\mathcal{V}(t)$, it can be simplified to Eq.~(\ref{HamMuStaNonHermComm}):
\begin{equation}\label{vonNeuStaNon}
\frac{\partial\tilde{\mu}_k^\dagger(0)}{\partial t}=-i\left[\hat{H}_{\rm rot}(t), \tilde{\mu}_k^\dagger(0)\right].
\end{equation}
Equations~(\ref{HamMuStaNonHermComm}) and (\ref{vonNeuStaNon}) hold only if $\hat{H}_{\rm rot}(t)$ is assumed in the form of Eq.~(\ref{HamMuStaNonHerm}). Any extra non-Hermitian component $\hat{\mu}_{k}(0)\hat{\mu}_{k'}(0)$ would invalidate this condition. Specifically, if Eq.~(\ref{HamMuStaNonHerm}) is modified as
\begin{equation}\label{HamMuStaNonExt}
\hat{H}_{\rm rot}'(t)=\hat{H}_{\rm rot}(t)+\mathcal{H}_{N+k,k'}(t)\hat{\mu}_{k}(0)\hat{\mu}_{k'}(0),
\end{equation}
then the commutation between $\hat{H}_{\rm rot}'(t)$ and $\tilde{\mu}_k^\dagger(0)$ becomes
\begin{equation}\label{HamMuStaNonHermCommExt}
-i\left[\hat{H}_{\rm rot}'(t),\tilde{\mu}_k^\dagger(0)\right]
=-ie^{-if_k(t)}\left[\dot{f}_k(t)\hat{\mu}_k^\dagger(0)
+\mathcal{H}_{N+k,k'}\hat{\mu}_k'(0)\right]
\neq\frac{\partial\tilde{\mu}_k^\dagger(0)}{\partial t}.
\end{equation}
The same failure occurs for other choices of $k'$ and $\tilde{\mu}_k^\dagger(0)$. Then the proof about the sufficient condition for obtaining $\hat{H}_{\rm rot}(t)$ in Eq.~(\ref{HamMuStaNonHerm}) is completed.

Using the Hamiltonian $\hat{H}_{\rm rot}(t)$ in Eq.~(\ref{HamMuStaNonHerm}), the time evolution operators in the ket and bra spaces can be formally expressed as $\hat{U}_{\rm rot}(t)=\hat{T}\exp[-i\int_0^t\hat{H}_{\rm rot}(s)ds]$ and $\hat{V}_{\rm rot}(t)=\hat{T}\exp[-i\int_0^t\hat{H}_{\rm rot}^\dagger(s)ds]$, respectively, with the time-ordering operator $\hat{T}$. By the non-Hermitian Heisenberg equation $\hat{\mathcal{O}}_H(t)=\hat{V}_{\rm rot}^\dagger(t)\hat{\mathcal{O}}_S\hat{U}_{\rm rot}(t)$~\cite{Miao2016Investigation}, the dynamics of the ancillary operator $\hat{\mu}_k^\dagger(0)$ is found to be
\begin{equation}\label{mu1Heisen}
\begin{aligned}
\frac{d\hat{v}_k^\dagger(t)}{dt}&=i\hat{V}_{\rm rot}^\dagger(t)\left[\hat{H}_{\rm rot}(t),\hat{\mu}_k^\dagger(0)\right]\hat{U}_{\rm rot}(t)=i\dot{f}_k(t)\hat{V}_{\rm rot}^\dagger(t)\hat{\mu}_k^\dagger(0)\hat{U}_{\rm rot}(t)=i\dot{f}_k(t)\hat{v}_k^\dagger(t),
\end{aligned}
\end{equation}
which yields
\begin{equation}\label{NonHermDynSol}
\hat{v}_k^\dagger(t)=e^{if_k(t)}\hat{v}_k^\dagger(0).
\end{equation}
Rotating back to the original picture via Eqs.~(\ref{Unitrans}) and (\ref{HamMuSta}), the dynamics of $\hat{\mu}_k^\dagger(0)$ is rewritten as
\begin{equation}\label{evolveNon}
\hat{\mu}_k^\dagger(0)\rightarrow\mathcal{V}(t)\hat{v}_k^\dagger(t)\mathcal{V}^\dagger(t)=e^{if_k(t)}\hat{\mu}_k^\dagger(t),
\end{equation}
which is the same as Eq.~(\ref{NonHermDynOri}).

\section{Derivation of non-Hermitian squeezing Hamiltonian}\label{HeffDerive}

This section presents detailed derivations of the non-Hermitian squeezing Hamiltonian for the single-mode and double-mode systems in Eqs.~(\ref{HamNHsing}) and (\ref{HamTwo}), respectively, from the adjoint Lindblad master equation. The quantum jump terms are fully retained throughout the derivations. Also, the consistency between the non-Hermitian Schr\"odinger equation and the Lindblad master equation is numerically confirmed in the single-mode case.

\subsection{Single-mode case}\label{HeffDerivesingle}\label{HeffDeriveSing}

The dynamics of an open single-mode bosonic system under a two-photon driving field can be described by the master equation as~\cite{Carmichael1999statistical}
\begin{equation}\label{masterone}
\frac{d}{dt}\rho=-i[\hat{H}_{\rm coh}(t),\rho]+\gamma\mathcal{L}[\hat{a}_1]\rho,
\end{equation}
where the coherent Hamiltonian $\hat{H}_{\rm coh}(t)=\omega\hat{a}_1^\dagger\hat{a}_1+[\Omega(t)\exp{(i\varphi)}(\hat{a}_1^\dagger)^2+{\rm H.c.}]$ provides the eigenenergy of the bosonic mode and the intensity of the driving field. The Lindblad superoperator $\mathcal{L}[\hat{a}_1]\rho\equiv\hat{a}_1\rho \hat{a}_1^\dagger-\{\hat{a}_1^\dagger\hat{a}_1,\rho\}/2$ describes the system dissipation  with a damping rate $\gamma$.

Under Eq.~(\ref{masterone}), the Schr\"odinger-picture operator $\hat{\mathcal{O}}_S$ is connected to the Heisenberg-picture operator $\hat{\mathcal{O}}_H(t)$ by
\begin{equation}\label{masterSHone}
{\rm Tr}\left[\hat{\mathcal{O}}_S\dot{\rho}(t)\right]={\rm Tr}\left[\hat{\mathcal{O}}_S\left(-i[\hat{H}_{\rm coh}(t),\rho]+\gamma\mathcal{L}[\hat{a}_1]\rho\right)\right]={\rm Tr}\left[\left(i[\hat{H}_{\rm coh}(t),\hat{\mathcal{O}}_S]+\gamma\mathcal{L}^\dagger[\hat{a}_1]\hat{\mathcal{O}}_S\right)
\rho\right]={\rm Tr}\left[\dot{\hat{\mathcal{O}}}_H(t)\rho(0)\right],
\end{equation}
with the Hermitian conjugate superoperator $\mathcal{L}^\dagger[\hat{a}_1]\mathcal{O}_S\equiv \hat{a}_1^\dagger\mathcal{O}_S\hat{a}_1-\{\hat{a}_1^\dagger \hat{a}_1,\mathcal{O}_S\}/2$. The second equivalence of Eq.~(\ref{masterSHone}) has used the cyclic property of trace. Accordingly, the dynamics of $\hat{\mathcal{O}}_H(t)$ can be described by the adjoint Lindblad master equation as
\begin{equation}\label{OHdynamicOne}
\frac{d}{dt}\hat{\mathcal{O}}_H(t)=i[\hat{H}_{\rm coh}(t),\hat{\mathcal{O}}_S]+\gamma\mathcal{L}^\dagger[\hat{a}_1]\hat{\mathcal{O}}_S.
\end{equation}

Using Eq.~(\ref{OHdynamicOne}), the dynamics of the modes $\hat{a}$ and $\hat{a}^\dagger$ can be expressed as
\begin{equation}\label{abdynamicOne}
\begin{aligned}
\frac{d}{dt}\hat{a}_1(t)&=-i\omega\hat{a}_1-\frac{\gamma}{2}\hat{a}_1-i2\Omega(t)e^{i\varphi}\hat{a}_1^\dagger,\\
\frac{d}{dt}\hat{a}_1^\dagger(t)&=i\omega\hat{a}_1^\dagger-\frac{\gamma}{2}\hat{a}_1^\dagger+i2\Omega(t)e^{-i\varphi}\hat{a}_1,
\end{aligned}
\end{equation}
which is equivalent to that governed by the non-Hermitian Hamiltonian in Eq.~(\ref{HamNHsing}).

\begin{figure}[htbp]
\centering
\includegraphics[width=0.5\linewidth]{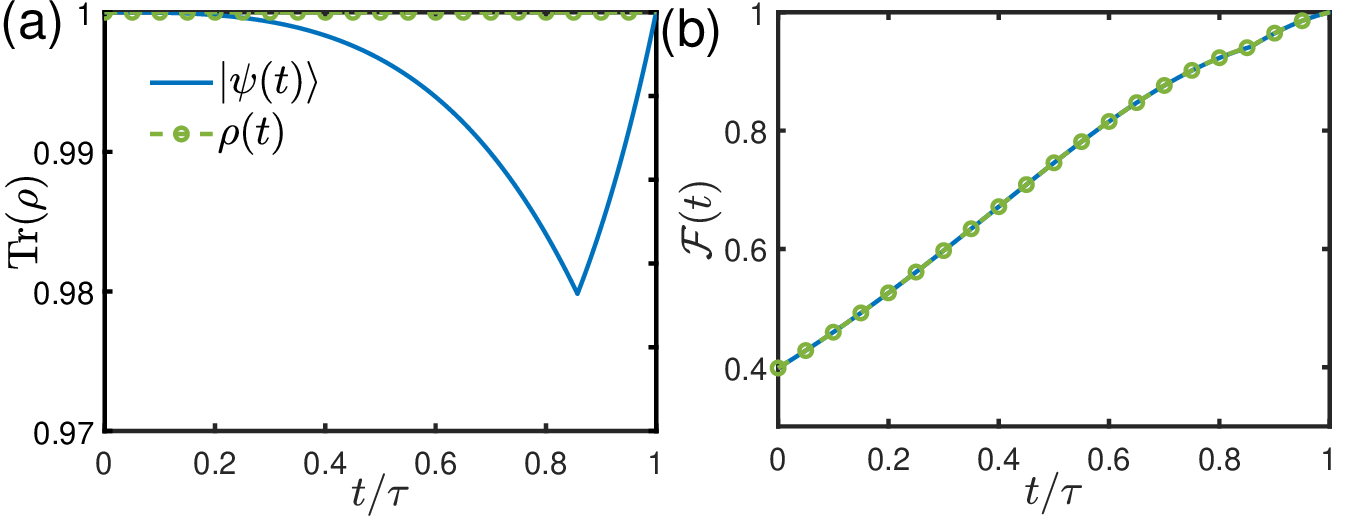}
\caption{Dynamics of (a) the traces of the density matrix obtained by either Schr\"odinger equation or Lindblad master equation and (b) the relevant fidelities with respect to the target squeezed state. $\lambda=10$ and $\tau/T=1$. The other parameters are the same as those for the case of $\tau/T=3/2$ in Fig.~\ref{onesqueez}.}\label{comparison}
\end{figure}

The equivalence between the effective non-Hermitian Hamiltonian Schr\"odinger equation and the full master equation can be further confirmed by the numerical simulation in Fig.~\ref{comparison}, where all the parameters for these two equations are identical. In Fig.~\ref{comparison}(a), it is found that the density-matrix trace ${\rm Tr}[|\psi(t)\rangle\langle\psi(t)|]$ obtained by the Schr\"odinger equation is conserved at both initial and target moments, i.e., ${\rm Tr}[|\psi(t)\rangle\langle\psi(t)|]=1$ when $t=0$ and $t=\tau$. In comparison, the density matrix $\rho(t)$ obtained by the Lindblad master equation is trace preserving during the whole evolution. In Fig.~\ref{comparison}(b), the fidelity dynamics by these two methods is found to be coincident with each other.

\subsection{double-mode case}\label{HeffDerivedouble}\label{HeffDeriveTwo}

Consider an open two-mode bosonic system coupled to two individual environments, the system dynamics can be described by the master equation as~\cite{Carmichael1999statistical}
\begin{equation}\label{master}
\frac{d}{dt}\rho=-i[\hat{H}_{\rm coh}(t),\rho]+\gamma_1\mathcal{L}[\hat{a}_1]\rho+\gamma_2\mathcal{L}[\hat{a}_2]\rho,
\end{equation}
where the coherent Hamiltonian $\hat{H}_{\rm coh}(t)=\omega_1\hat{a}_1^\dagger \hat{a}_1+\omega_2\hat{a}_2^\dagger \hat{a}_2+[g(t)\exp(i\varphi)\hat{a}_1^\dagger \hat{a}_2^\dagger+{\rm H.c.}]$ describes the eigenenergies of the two modes and the squeezing interaction between them. The Lindblad superoperators $\mathcal{L}[\hat{a}_1]\rho$ and $\mathcal{L}[\hat{a}_2]\rho$ are associated with the local damping rates $\gamma_1$ and $\gamma_2$, respectively.

Similar to Eq.~(\ref{masterSHone}), the Schr\"odinger-picture operator $\hat{\mathcal{O}}_S$ governed by Eq.~(\ref{master}) is connected to the Heisenberg-picture operator $\hat{\mathcal{O}}_H(t)$ by
\begin{equation}\label{masterSH}
\begin{aligned}
&{\rm Tr}\left[\hat{\mathcal{O}}_S\dot{\rho}(t)\right]={\rm Tr}\left[\hat{\mathcal{O}}_S\left(-i[\hat{H}_{\rm coh}(t),\rho]+\gamma_1\mathcal{L}[\hat{a}_1]\rho+\gamma_2\mathcal{L}[\hat{a}_2]\rho\right)\right]\\
&={\rm Tr}\left[\left(i[\hat{H}_{\rm coh}(t),\hat{\mathcal{O}}_S]+\gamma_1\mathcal{L}^\dagger[\hat{a}_1]\hat{\mathcal{O}}_S
+\gamma_2\mathcal{L}^\dagger[\hat{a}_2]\hat{\mathcal{O}}_S\right)\rho\right]={\rm Tr}\left[\dot{\hat{\mathcal{O}}}_H(t)\rho(0)\right].
\end{aligned}
\end{equation}
Accordingly, the time derivative of $\hat{\mathcal{O}}_H(t)$ can be expressed by the adjoint Lindblad master equation as
\begin{equation}\label{OHdynamic}
\frac{d}{dt}\hat{\mathcal{O}}_H(t)=i[\hat{H}_{\rm coh}(t),\hat{\mathcal{O}}_S]+\gamma_1\mathcal{L}^\dagger[\hat{a}_1]\hat{\mathcal{O}}_S
+\gamma_2\mathcal{L}^\dagger[\hat{a}_2]\hat{\mathcal{O}}_S.
\end{equation}
Using Eq.~(\ref{OHdynamic}), the dynamics about the modes $\hat{a}_1$ and $\hat{a}_2$ can be obtained as
\begin{equation}\label{abdynamic}
\begin{aligned}
\frac{d}{dt}\hat{a}_1(t)&=-i\omega_1\hat{a}_1-\frac{\gamma_1}{2}\hat{a}_1-ige^{i\varphi}\hat{a}_2^\dagger,\\
\frac{d}{dt}\hat{a}_2(t)&=-i\omega_2\hat{a}_2-\frac{\gamma_2}{2}\hat{a}_2-ige^{i\varphi}\hat{a}_1^\dagger.
\end{aligned}
\end{equation}
They are the same as the Heisenberg equations governed by the non-Hermitian Hamiltonian in Eq.~(\ref{HamTwo}).

\section{Full symplectic matrix for two-modes system and a brief recipe for general ancillary operators}\label{recipe}

This section first introduces the completed symplectic matrix to construct the ancillary operators for two-mode bosonic systems with squeezing interactions, and then provides a brief recipe about constructing the ancillary operators for a general $N$-mode bosonic system.

For the two-mode bosonic system governed by the squeezing Hamiltonian~(\ref{HamTwo}), we construct the time-dependent ancillary operators by a $4\times 4$ symplectic matrix according to Eq.~(\ref{TransM}):
\begin{equation}\label{AnciFull}
\begin{pmatrix}\hat{\mu}_1(t)\\ \hat{\mu}_2(t)\\ \hat{\mu}_1^\dagger(t)\\ \hat{\mu}_2^\dagger(t)\end{pmatrix}=\begin{pmatrix}\cosh\theta(t)&0&0&-\sinh\theta(t)e^{-i\alpha(t)}\\
0&\cosh\theta(t)&-\sinh\theta(t)e^{-i\alpha(t)}&0\\0&-\sinh\theta(t)e^{i\alpha(t)}&\cosh\theta(t)&0\\
-\sinh\theta(t)e^{i\alpha(t)}&0&0&\cosh\theta(t)\end{pmatrix}\begin{pmatrix}\hat{a}_1\\ \hat{a}_2\\ \hat{a}_1^\dagger\\ \hat{a}_2^\dagger\end{pmatrix}.
\end{equation}
Due to the symmetric structure of the symplectic matrix, Eq.~(\ref{AnciFull}) can be reduced to $[\hat{\mu}_1(t),\hat{\mu}_2^\dagger(t)]^T=\mathcal{M}(t)(\hat{a}_1,\hat{a}_2^\dagger)^T$ with $\mathcal{M}(t)$ given in Eq.~(\ref{Mone}) as employed in the main text.

In the following, we outline a brief recipe about constructing the ancillary operators for $N$-mode bosonic systems with squeezing interactions described by the Hamiltonian~(\ref{HamMuAppend}). The construction order is suggested as (1) $\{\hat{a}_1,\hat{a}_2^\dagger\}\rightarrow\{\hat{\mu}_1(t),\hat{b}_1^\dagger(t)\}$, $\{\hat{b}_1(t),\hat{a}_3^\dagger\}\rightarrow\{\hat{\mu}_2(t),\hat{b}_2^\dagger(t)\}$, $\{\hat{b}_2(t),\hat{a}_4^\dagger\}\rightarrow\{\hat{\mu}_3(t),\hat{b}_3^\dagger(t)\}$, $\cdots$, $\{\hat{b}_{N-3}(t),\hat{a}_{N-1}^\dagger\}\rightarrow\{\hat{\mu}_{N-2}(t),\hat{b}_{N-2}^\dagger(t)\}$ and (2) $\{\hat{b}_{N-2}(t),\hat{a}_N^\dagger\}\rightarrow\{\hat{\mu}_{N-1}(t),\hat{\mu}_N^\dagger(t)\}$. It is straightforward to verify that the time-dependent ancillary and bright operators satisfy the canonical commutation relations as $[\hat{\mu}_j(t), \hat{\mu}_k^\dagger(t)]=\delta_{jk}$ and $[\hat{b}_j(t),\hat{b}_k^\dagger(t)]=\delta_{jk}$, while they commute as $[\hat{\mu}_j(t), \hat{b}_k(t)]=0$.

Using a SU(1,1)-like rotating matrix, the ancillary and bright operators in Step (1) can be formulated as
\begin{equation}\label{Ancibrig}
\begin{aligned}
&\left(\begin{array}{c} \hat{\mu}_k(t) \\ \hat{b}_k^\dagger(t) \end{array} \right)=\left(\begin{array}{cc}
\cosh\theta_k(t) & -\sinh\theta_k(t)e^{-i\alpha_k(t)} \\ -\sinh\theta_k(t)e^{i\alpha_k(t)} & \cosh\theta_k(t)
\end{array}\right)\left(\begin{array}{c}\hat{b}_{k-1}(t)\\ \hat{a}_{k+1}^\dagger \end{array} \right),
\end{aligned}
\end{equation}
where $k=1,2,\cdots,N-2$ and $\hat{b}_0(t)\equiv\hat{a}_1$. Step (2) can be expressed as
\begin{equation}\label{Ancibrig2}
\begin{aligned}
&\left(\begin{array}{c} \hat{\mu}_{N-1}(t) \\ \hat{\mu}_N^\dagger(t) \end{array} \right)=\left(\begin{array}{cc}
\cosh\theta_{N-1}(t) & -\sinh\theta_{N-1}(t)e^{-i\alpha_{N-1}(t)} \\ -\sinh\theta_{N-1}(t)e^{i\alpha_{N-1}(t)} & \cosh\theta_{N-1}(t)
\end{array}\right)\left(\begin{array}{c}\hat{b}_{N-2}(t)\\ \hat{a}_N^\dagger \end{array} \right).
\end{aligned}
\end{equation}
Here the time-dependent parameters $\theta_k(t)$ and $\alpha_k(t)$, with $k=1,2,\cdots,N-1$, are relative to the squeezing strength and phase, respectively. The ancillary operators defined in Eqs.~(\ref{Ancibrig}) and (\ref{Ancibrig2}) can be compactly expressed as
\begin{equation}\label{TimeAnci}
\vec{\mu}_t^T=\mathcal{M}(t)\vec{a}^T, \quad \vec{\mu}_t\equiv\left[\hat{\mu}_1(t),\hat{\mu}_2(t), \cdots,\hat{\mu}_{N-1}(t),\hat{\mu}_N^\dagger(t)\right], \quad \vec{a}\equiv\left[\hat{a}_1,\hat{a}_2^\dagger,\hat{a}_3^\dagger,\cdots,\hat{a}_N^\dagger\right],
\end{equation}
where the $N\times N$ unitary transformation matrix $\mathcal{M}(t)$ admits a general representation of the form
\begin{equation}\label{unitary}
\mathcal{M}(t)=\left[\vec{M}_1(t), \vec{M}_2(t), \cdots, \vec{M}_N(t)\right]^T
\end{equation}
with the row vectors of $N$ dimensionality
\begin{equation}\label{unitarVec}
\begin{aligned}
\vec{M}_1(t)&=\left(\cosh\theta_1(t),-\sinh\theta_1(t)e^{-i\alpha_1(t)},0,\cdots,0\right),\\
\vec{M}_k(t)&=\left[\cosh\theta_k(t)\vec{b}_{k-1}(t),-\sinh\theta_k(t)e^{-i\alpha_k(t)},0,\cdots,0\right],\\
\vec{M}_{N-1}(t)&=\left[\cosh\theta_{N-1}(t)\vec{b}_{N-2}(t),-\sinh\theta_{N-1}(t)e^{-i\alpha_{N-1}(t)}\right],\\
\vec{M}_N(t)&=\vec{b}_{N-1}(t),
\end{aligned}
\end{equation}
where $k$ runs from $2$ to $N-2$. The bright vector $\vec{b}_k(t)$ is a row vector of $k+1$ dimensionality defined as
\begin{equation}\label{unitarVecBri}
\vec{b}_k(t)\equiv\left[-\sinh\theta_k(t)e^{i\alpha_k(t)}\vec{b}_{k-1}(t), \cosh\theta_k(t)\right]
\end{equation}
where $1\leq k\leq N-1$ and $\vec{b}_0(t)\equiv 1$. Using Eq.~(\ref{unitarVecBri}), the bright operators formulated in Eqs.~(\ref{Ancibrig}) and (\ref{Ancibrig2}) can be expressed as inner products $\hat{b}_k^\dagger(t)=\vec{b}_k(t)\cdot\vec{a}_{k+1}^T$ with $\vec{a}_k^T=(\hat{a}_1,\hat{a}_2^\dagger,\cdots,\hat{a}_k^\dagger)^T$ and $\hat{b}_{N-1}^\dagger(t)\equiv\hat{\mu}_N(t)$, e.g., $\hat{b}_1^\dagger(t)=-\sinh\theta_1(t)e^{i\alpha_1(t)}\hat{a}_1+\cosh\theta_1(t)\hat{a}_2^\dagger$.

For example, one can consider a three-modes bosonic system governed by the Hamiltonian~\cite{McDonald2018Phase,Busnaina2024Quantum}
\begin{equation}\label{HamthreeI}
\hat{H}(t)=g_1(t)e^{i\varphi_1}\hat{a}_1\hat{a}_3^\dagger+g_2(t)e^{i\varphi_2}\hat{a}_2^\dagger\hat{a}_3^\dagger+{\rm H.c.}
\end{equation}
where $\varphi_1$ and $\varphi_2$ denote the phases, and $g_1(t)$ and $g_2(t)$ represent the exchange coupling strength and the squeezing coupling strength, respectively. For such bosonic systems, the time-dependent ancillary operators can be constructed in accordance to Eqs.~(\ref{unitary}) and (\ref{unitarVec}) as
\begin{equation}\label{Ancithree}
\begin{aligned}
\hat{\mu}_1(t)&=\cosh\theta_1(t)\hat{a}_1-\sinh\theta_1(t)e^{-i\alpha_1(t)}\hat{a}_2^\dagger,\\
\hat{\mu}_2(t)&=\cosh\theta_2(t)\hat{b}_1-\sinh\theta_2(t)e^{-i\alpha_2(t)}\hat{a}_3^\dagger,\\
\hat{\mu}_3^\dagger(t)&=-\sinh\theta_2(t)e^{i\alpha_2(t)}\hat{b}_1+\cosh\theta_2(t)\hat{a}_3^\dagger
\end{aligned}
\end{equation}
with the bright mode $\hat{b}_1=-\sinh\theta_1(t)e^{-i\alpha_1(t)}\hat{a}_1^\dagger+\cosh\theta_1(t)\hat{a}_2$. The unitary transformation $\mathcal{V}(t)$ that maps the time-dependent ancillary operators to their time-independent counterparts can be expressed as
\begin{equation}\label{UnitarThree}
\mathcal{V}(t)=S_1^\dagger\left[\theta_1(t)e^{-i\alpha_1(t)}\right]S_1\left[\theta_1(0)e^{-i\alpha_1(0)}\right]
S_2^\dagger\left[\theta_2(t)e^{-i\alpha_2(t)}\right]S_2\left[\theta_2(0)e^{-i\alpha_2(0)}\right]
\end{equation}
with the two-mode and three-mode squeezing operators defined as
\begin{equation}\label{ThreeSqueez}
S_1(\xi_1)\equiv e^{\xi_1^*\hat{a}_1\hat{a}_2-\xi\hat{a}_1^\dagger\hat{a}_2^\dagger},\quad S_2(\xi_2)\equiv e^{\xi_2^*\hat{b}_1\hat{a}_3-\xi\hat{b}_1^\dagger\hat{a}_3^\dagger},
\end{equation}
respectively.

Substituting the ancillary operators in Eq.~(\ref{Ancithree}) into the Heisenberg equation~(\ref{vonNeu}) with $\hat{H}(t)$ in Eq.~(\ref{HamthreeI}), one can obtain the constraints on the phases, the exchange coupling strength, and the squeezing coupling strengths:
\begin{equation}\label{constrainThree}
\begin{aligned}
\varphi_1&=\varphi_2-\alpha_1=\frac{\pi}{2},\quad \alpha_2(t)=0,\\
g_1(t)&=\dot{\theta}_2(t)\sinh\theta_1,\quad g_2(t)=-\dot{\theta}_2(t)\cosh\theta_1.
\end{aligned}
\end{equation}
Under these conditions, $\hat{\mu}_1(t)$ becomes a decoupled mode since $[\hat{H}(t), \hat{\mu}_1(t)]=0$, while $\hat{\mu}_2(t)$ and $\hat{\mu}_3^\dagger(t)$ can be activated as nonadiabatic Heisenberg passages.

\end{document}